\documentclass[prb,twocolumn,superscriptaddress]{revtex4-2}
\usepackage{graphicx,empheq,amssymb,scalerel,tikz}
\usepackage{xcolor}
\usepackage[normalem]{ulem}
\usepackage[colorlinks=true, citecolor=blue, urlcolor=blue, linkcolor=blue ]{hyperref}
\usetikzlibrary{svg.path}
\definecolor{orcidlogocol}{HTML}{A6CE39}
\tikzset{
	orcidlogo/.pic={
		\fill[orcidlogocol] svg{M256,128c0,70.7-57.3,128-128,128C57.3,256,0,198.7,0,128C0,57.3,57.3,0,128,0C198.7,0,256,57.3,256,128z};
		\fill[white] svg{M86.3,186.2H70.9V79.1h15.4v48.4V186.2z}
		svg{M108.9,79.1h41.6c39.6,0,57,28.3,57,53.6c0,27.5-21.5,53.6-56.8,53.6h-41.8V79.1z M124.3,172.4h24.5c34.9,0,42.9-26.5,42.9-39.7c0-21.5-13.7-39.7-43.7-39.7h-23.7V172.4z}
		svg{M88.7,56.8c0,5.5-4.5,10.1-10.1,10.1c-5.6,0-10.1-4.6-10.1-10.1c0-5.6,4.5-10.1,10.1-10.1C84.2,46.7,88.7,51.3,88.7,56.8z};}}
\newcommand\orcid[1]{\href{https://orcid.org/#1}{\mbox{\scalerel*{\begin{tikzpicture}[yscale=-1,transform shape]\pic{orcidlogo};\end{tikzpicture}}{|}}}}

\begin{document}
\title{Topological semimetal with coexisting nodal points and nodal lines}
\author{Bing-Bing Luo}
\affiliation{Lanzhou Center for Theoretical Physics, Key Laboratory of Theoretical Physics of Gansu Province, Key Laboratory of Quantum Theory and Applications of MoE, Gansu Provincial Research Center for Basic Disciplines of Quantum Physics, Lanzhou University, Lanzhou 730000, China}
\author{Ming-Jian Gao\orcid{0000-0002-6128-8381}}
\affiliation{Lanzhou Center for Theoretical Physics, Key Laboratory of Theoretical Physics of Gansu Province, Key Laboratory of Quantum Theory and Applications of MoE, Gansu Provincial Research Center for Basic Disciplines of Quantum Physics, Lanzhou University, Lanzhou 730000, China}
\author{Jun-Hong An\orcid{0000-0002-3475-0729}}
\email{anjhong@lzu.edu.cn}
\affiliation{Lanzhou Center for Theoretical Physics, Key Laboratory of Theoretical Physics of Gansu Province, Key Laboratory of Quantum Theory and Applications of MoE, Gansu Provincial Research Center for Basic Disciplines of Quantum Physics, Lanzhou University, Lanzhou 730000, China}
\begin{abstract}
Featuring exotic quantum transport and surface states, topological semimetals can be classified into nodal-point, nodal-line, and nodal-surface semimetals according to the degeneracy and dimensionality of their nodes.  However, a topological semimetal that possesses both nodal points and nodal lines is rarely reported. Here, we propose a scheme to construct this type of topological semimetal, which simultaneously exhibits hinge Fermi arcs and drumhead surface states. Then, by applying periodic driving on the system, we find a hybrid-order topological semimetal with nodal points and rich nodal-line structures and its conversion into a first-order topological semimetal, which are absent in a static system. Our results enrich the family of topological semimetals, and establish a foundation for further exploration of their potential applications.
\end{abstract}
\maketitle

\section{Introduction}
The topological phases of matter, which are broadly classified as insulators and semimetals according to energy band gaps, have emerged as a leading research domain in condensed matter physics \cite{RevModPhys.82.3045,RevModPhys.88.035005,RevModPhys.83.1057,RevModPhys.90.015001,RevModPhys.93.025002}. The topological insulator has been extended from the traditional topological energy band theory to a higher-order one \cite{Benalcazar.2017,Xie.2021,PhysRevLett.119.246401,PhysRevLett.122.204301,Ghosh_2024}. Recently, significant research attention has focused on topological semimetals. Depending on the symmetry endowed degeneracy and dimensionality of their nodes, topological semimetals can be classified as Dirac \cite{PhysRevLett.108.140405,PhysRevLett.115.126803,PhysRevLett.119.026404,Z.K.Liu.2014,PhysRevLett.113.027603,Wieder.Benjamin.J.2020}, Weyl \cite{Su-Yang.Xu.Ilya-Belopolski.2015,PhysRevX.5.031013,Yang.L.X.2015,PhysRevLett.125.146401,PhysRevLett.125.266804,Luo.Li.2021,PhysRevB.109.054201,PhysRevB.106.045424}, nodal-line \cite{PhysRevB.84.235126,PhysRevB.96.041103,PhysRevB.97.161111,PhysRevLett.125.126403}, and nodal-surface \cite{PhysRevLett.132.186601} semimetals. In contrast to Dirac and Weyl semimetals, which feature discrete points at the intersection of the conduction and valence bands, nodal-line and -surface semimetals are distinguished by the formation of a one-dimensional line and two-dimensional (2D) surface at the intersection of the energy bands, respectively.

It has been demonstrated that distinct topological semimetals can be converted into each other. Phase transitions from nodal-line and Dirac semimetals to a Weyl one have been created by breaking mirror symmetry \cite{PhysRevB.104.235136} and introducing circularly polarized light \cite{PhysRevLett.117.087402,PhysRevB.96.041206,PhysRevB.107.L121407}. An addition of a spin-orbit coupling has been shown to convert a nodal-line semimetal into a Dirac one \cite{Fang_2016,PhysRevLett.116.127202}. By introducing a $\mathcal{PT}$-invariant perturbation, each Dirac point is expanded into a nodal ring that separates the hinge Fermi arcs, which signifies a second-order nodal-line semimetal \cite{PhysRevLett.125.126403}. However, the topological semimetals in previous studies possess only one type, either a nodal-line semimetal or a nodal-point semimetal. From an application perspective, the transport properties of different orders and types of topological semimetals are generally different \cite{PhysRevLett.122.196603,PhysRevLett.124.056402}. A nodal-point semimetal not only induces a negative magnetoresistance due to the chiral anomaly \cite{PhysRevB.88.104412,PhysRevB.95.161306,PhysRevB.86.115133,PhysRevLett.111.027201,PhysRevX.5.031023,PhysRevB.89.085126}, but also exhibits high carrier mobility enabling an ultra-low-power electronic transport \cite{LiangTian2015}. A nodal-line semimetal exhibits a heightened sensitivity to surface perturbations and strain and leads to a strong anisotropy in the optical and transport properties due to its drumhead surface states \cite{PhysRevLett.121.166802,PhysRevLett.120.146602,Hu2024}. The nodal-line semimetals realized in photonic crystal platforms have been used to design optical waveguides with topologically protected edge states \cite{Hu2024}. They can effectively suppress backscattering and provide the basis for high-sensitivity sensors and optical switches. The photonic Weyl metamaterial provides a new way to generate vector and vortex beams \cite{PhysRevLett.125.093904}. 
In circuit systems, a topological semimetal can be used to achieve robust signal transmission with backscattering suppression \cite{YANG20241,PhysRevResearch.3.023056,Rafi-Ul-Islam2020}.
It is expected that topological semimetals with coexisting nodal points and lines may simultaneously exploit these peculiar features to design multifunctional devices. Thus, it is necessary to find a topological semimetal that inherits both nodal lines and points, as well as both first
and second orders. Although hybrid-order topological insulators and superconductors have been widely studied \cite{PhysRevLett.126.156801,PhysRevB.109.115422,PhysRevB.107.235132,PhysRevApplied.21.044002,HuangShengJie2024,PhysRevB.108.125125,10.1063/5.0238775,PhysRevB.109.134302,PhysRevB.111.195107}, a hybrid-order topological semimetal with coexisting nodal points and lines has seldom been reported.  

\begin{figure}[tbp]
\centering
\includegraphics[width=\columnwidth]{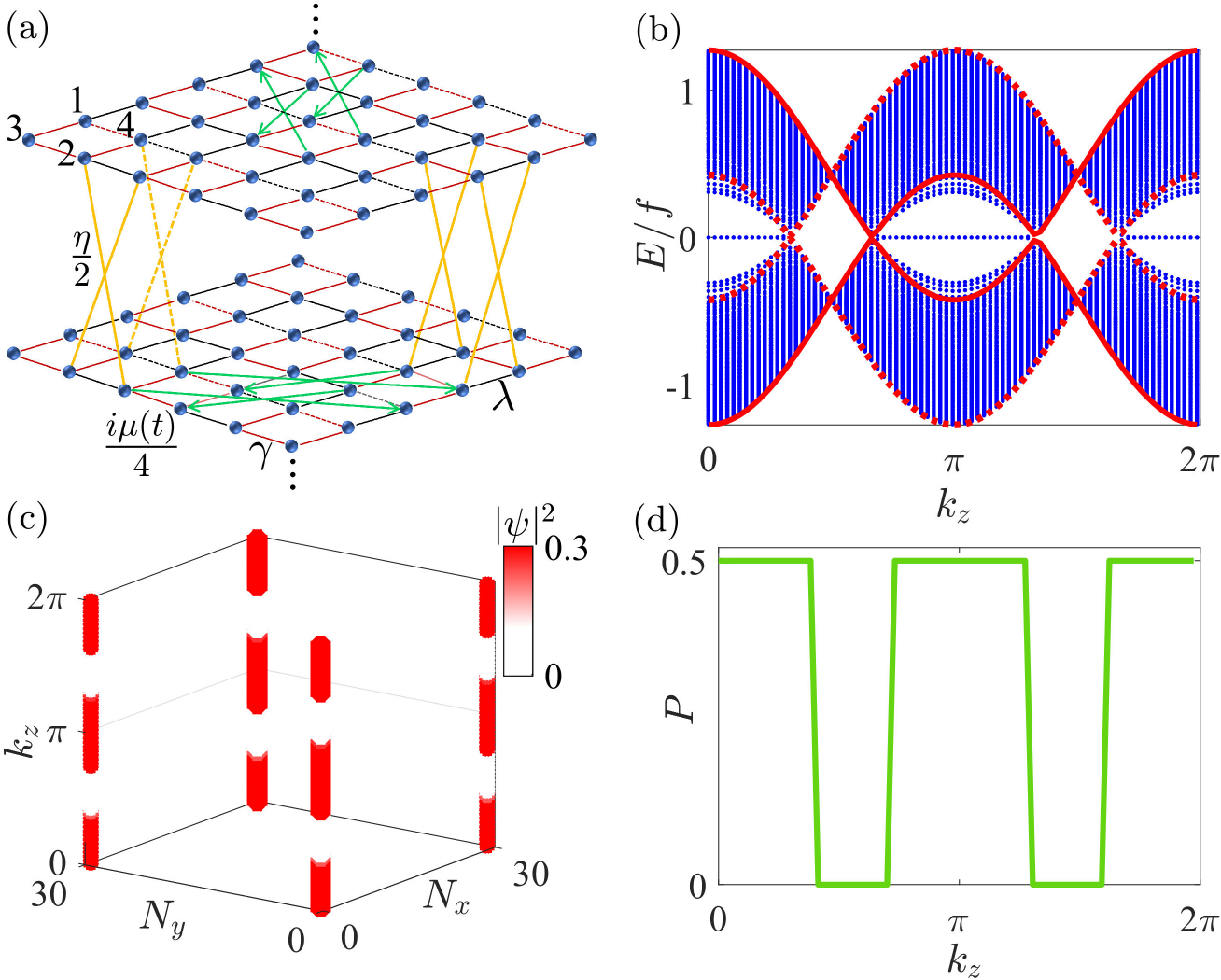}
\caption{(a) Schematic diagram of a lattice configuration with intracell hopping $\gamma$, intercell hopping $\lambda$, interlayer hopping $\eta/2$, and diagonal intercell hopping $i\mu/4$ (green lines along the arrow direction) with inverse hopping $-i\mu/4$ (green lines inverse to the arrow direction), respectively. The dashed lines denote the hopping rate with a $\pi$-phase difference from their corresponding solid lines. (b) Energy spectrum (blue dots) and dispersion relation along the high-symmetry line $k=0$ ($\pi$) in red solid (dashed) lines. (c) Hinge Fermi arcs  with the number of cells $N_{x}=N_{y}=30$. (d) Quadrupole moment $P$ changes with $k_{z}$. We use $\lambda=\mu=0$, $\gamma=0.3f$, and $\eta=0.6f$. $f$ is an energy scale to make the hopping rates dimensionless.}\label{fig:1}
\end{figure} 

Inspired by the fact that the coherent control by periodic driving called Floquet engineering has become a powerful tool to create novel topological phases \cite{PhysRevLett.121.036401,PhysRevLett.124.057001,PhysRevB.103.115308,PhysRevB.104.205117,PhysRevB.105.224312,PhysRevB.96.041126,Hübener.2017,PhysRevB.100.134301}, we propose a scheme to generate a hybrid-order topological semimetal with coexisting nodal points and nodal lines in a three-dimensional (3D) system. First, we find that a second-order Dirac semimetal can be converted into a second-order topological semimetal with coexisting nodal points and nodal lines by adding a hopping term that breaks the time-reversal symmetry. Second, such an exotic topological matter is further changed into either a hybrid-order or a purely first-order topological semimetal with coexisting nodal points and rich nodal lines by Floquet engineering. Our results enrich the family of topological semimetals and supply an extra control dimension for exploring their applications in designing multifunctional devices.

The paper is organized as follows. A static system supporting a second-order topological semimetal with coexisting nodal points and lines and its topological characterization are given in Sec. \ref{stcsst}. A hybrid-order topological semimetal with nodal points and rich nodal-line structures and its conversion to a purely first-order topological semimetal under Floquet engineering are studied in Sec. \ref{fenggd}. Finally, a discussion and a summary are given in Sec. \ref{ddccl}.

\section{Static system}\label{stcsst}
We investigate a system of spinless fermions moving on a 3D lattice [see Fig. \ref{fig:1}(a)]. Its momentum-space Hamiltonian reads $\hat{H} = \sum_{\textbf{k}}\hat{C}^{\dagger}_{\textbf{k}}\mathcal{H}$(\textbf{k})$\hat{C}_{\textbf{k}}$ with $\hat{C}^{\dagger}_{\textbf{k}}$ = ($\hat{C}^{\dagger}_{\textbf{k},1}$ $\hat{C}^{\dagger}_{\textbf{k},2}$ $\hat{C}^{\dagger}_{\textbf{k},3}$ $\hat{C}^{\dagger}_{\textbf{k},4}$) and
\begin{equation} \label{text}
\mathcal{H}(\textbf{k})=h_{1}\Gamma_{5}+h_{2}\Gamma_{3}+h_{3}\Gamma_{2}+h_{4}\Gamma_{1}+h_{5}\Gamma_{4},
\end{equation}
where $h_{1}=\gamma+\chi(k_{z})\textrm{cos}k_{x}$, $h_{2}=-\chi(k_{z})\textrm{sin}k_{x}$, $h_{3}=\gamma+\chi(k_{z})\textrm{cos}k_{y}$, $h_{4}=\chi(k_{z})\textrm{sin}k_{y}$, $h_{5}=\mu\sin k_x \sin k_y$, and $\chi(k_{z})=\lambda+\eta\cos k_{z}$. Each unit cell has four sublattices. $\gamma$ is the intracell hopping rate, $\lambda$ is the intercell hopping rate, $\eta$ is the interlayer hopping rate, $\mu$ is the nearest-neighbor diagonal intercell hopping rate, $\Gamma_{1,2,3}=\tau_{2}\sigma_{1,2,3}$, $\Gamma_{4}=\tau_{1}\sigma_{2}$, and $\Gamma_{5}=\tau_{1}\sigma_{0}$, where $\tau_{i}$ and $\sigma_{i}$ are Pauli matrices, and $\tau_{0}$ and $\sigma_{0}$ are identity matrices.

\begin{figure*}[!ht]
\centering
\includegraphics[width=2.0\columnwidth,trim=0.65cm 0cm 0cm 0cm,clip=false]{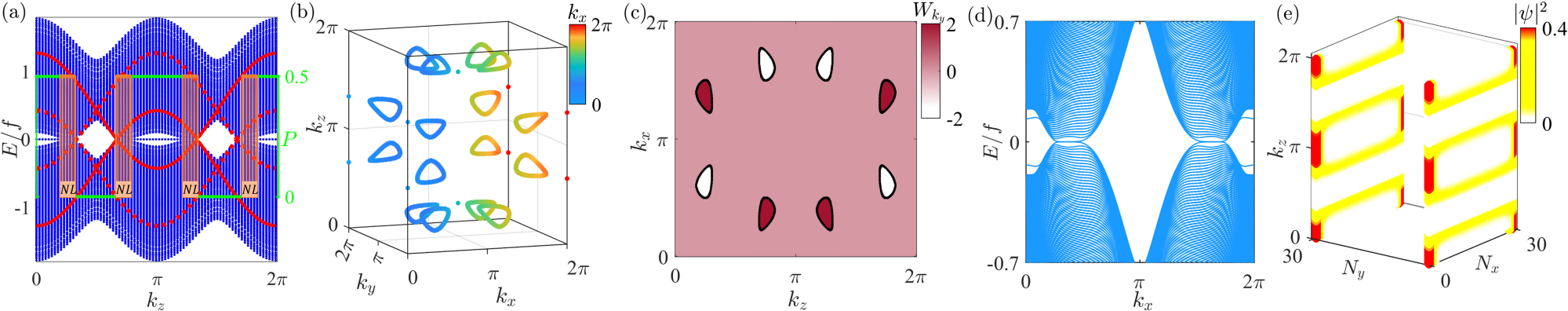}
\caption{(a) Energy spectrum (blue dots), quadrupole moment $P$ (green lines), and dispersion relation along the high-symmetry line $k = 0$ ($\pi$) in red solid (dashed) lines. The nodal line areas (NL) marked by NL are shaded by orange. (b) Distribution of nodal point and nodal line in the first Brillouin zone. (c) Winding number $W_{k_{y}}$ and nodal lines (black lines) in the $k_x$-$k_z$ plane.  (d) Dispersion relation under the condition of the $y$ direction open boundary and $x$ direction closed boundary reveals two flat bands at zero energy when $k_x\in [0.27,0.47]\pi$ and $[1.73,1.52]\pi$, which signify the drumhead surface states. (e) Coexistence of the hinge (red dots) and surface  (yellow dots) Fermi arcs. We use $\gamma=0.3f$, $\eta=0.6f$, $\lambda=0$, $\mu=0.85f$.}\label{fig:2}
\end{figure*}
The system in the absence of $\mu$ is a 3D generalization of the Benalcazar-Bernevig-Hughes model and hosts a second-order Dirac semimetal \cite{Benalcazar.2017}. It has the time-reversal $\mathcal{T}=K$, with $K$ being the complex conjugation, the spatial inversion $\mathcal{P}=\tau_0\sigma_y$, the mirror rotations under $\mathcal{M}_{x}$ = $\tau_{x}\sigma_{z}$, $\mathcal{M}_{y}$ = $\tau_{x}\sigma_{x}$, and $\mathcal{M}_{xy}=[(\tau_0-\tau_z)\sigma_x-(\tau_0+\tau_z)\sigma_z]/2$, and chiral $\mathcal{S}=\tau_z\sigma_0$ symmetries. The 3D topological semimetal can be seen as a stacking of 2D sliced systems parameterized by $k_{z}$. The topological phases of 2D sliced systems are well described by $\mathcal{H}(k,k,k_z)$ along the high-symmetry line $k_x=k_y\equiv k$. The dispersion relation is $E=\pm(\sum_{j=1}^{4}h_{j}^{2})^{\frac{1}{2}}$, which indicates that both the valence and conduction bands are doubly degenerate. Dirac points satisfying  $k=0$ or $\pi$ arise when $\left | \gamma  \right | =\left | \chi \right |$, at which a topological phase transition occurs in the 2D sliced systems \cite{Benalcazar.2017}. In our 3D system, $\chi$ becomes $\chi(k_z)=\lambda+\eta\cos k_z$. Therefore, we obtain the parameter boundaries of the presence of the hinge Fermi arcs at $k_z=\arccos[-(\lambda\pm\gamma)/\eta]$. The energy spectrum under the open $x,y$-boundary condition in Fig. \ref{fig:1}(b) verifies that gapped zero-mode states are present at these parameter boundaries. Their probability distributions in Fig. \ref{fig:1}(c) confirm the corner nature of the zero-mode states and verify the formation of a $k_z$-parametrized 2D sliced second-order topological insulator. Therefore, the stacking of these corner states along the $z$ direction forms the hinge Fermi arcs of a 3D second-order Dirac semimetal. The second-order corner states are characterized by the $k_{z}$-dependent quadrupole moment \cite{PhysRevB.103.L041115,PhysRevB.100.245135}
\begin{equation}
P=\Big[\frac{\text{Im}\ln\det\Theta}{2\pi}-\sum_{{\bf n},i;{\bf m},j} \frac{\Omega_{{\bf n},i;{\bf m},j}}{2N_xN_y}\Big]\text{mod}~1.\label{polrz}
\end{equation}
The elements of $\Theta$ are $\Theta_{\alpha\beta}\equiv \langle \psi_\alpha|e^{i2\pi \Omega/(N_xN_y)}|\psi_\beta\rangle$, $|\psi_\alpha\rangle$ satisfies $\hat{H}|\psi_\alpha \rangle=E_{\alpha}|\psi_\alpha\rangle$, and $\Omega_{{\bf n},i;{\bf m},j}=n_xn_y\delta_{{\bf n}{\bf m}}\delta_{ij}$ with $i,j=1,\cdots,4$ being the sublattices and $n_{x,y}$ being the number of unit cells. $P = 0.5$ successfully signifies the existence of corner states [see Fig. \ref{fig:1}(d)].

To generate a second-order topological semimetal with coexisting nodal points and nodal lines, we introduce a diagonal intercell hopping term $\mu$ in $h_{5}\Gamma_4$. It is easy to prove that the time-reversal $\mathcal{T}$ and mirror-rotation $\mathcal{M}_{xy}$ symmetries are broken, and the spatial inversion $\mathcal{P}$, mirror-rotation $\mathcal{M}_{x}$ and $\mathcal{M}_{y}$, and chiral $\mathcal{S}$ symmetries are still respected. Satisfied only by the term proportional to $\Gamma_4$, this symmetry requirement forms a sufficient condition to create the coexisting nodal points and lines. By diagonalizing $\mathcal{H}(\textbf{k})$, we obtain the dispersion relation $E=\pm\sqrt{\sum_{j=1}^{5}h_{j}^{2}\pm2[(h_{1}^{2}+h_{2}^{2}+h_{4}^{2})h_{5}^{2}]^{\frac{1}{2}}}$. The former doubly degenerate valence and conduction bands are split by $\mu$ contained in $h_5$. The energy spectrum under the open boundary condition in Fig. \ref{fig:2}(a) shows that the nodal points along the high-symmetry lines $k=0$ and $\pi$ in Fig. \ref{fig:1}(b) are preserved and the $k_z$-dependent 2D sliced second-order topological phases are well described by $P=0.5$. It is interesting to find that the energy bands are closed in several segments of $k_z$, which form the nodal lines [see Fig. \ref{fig:2}(b)]. It clearly signifies the coexistence of the nodal points and nodal lines. The regions enclosed by the nodal lines are first-order surface flat bands, which are described by the winding number \cite{RevModPhys.88.035005,PhysRevLett.123.246801}
\begin{equation}\label{Wky}
\begin{aligned}
W_{k_y}=\frac{1}{4\pi i}\int_{-\pi}^{\pi} \textrm{Tr}[\mathcal{S}\mathcal{F}(\textbf{k})\partial_{k_y} \mathcal{F}(\textbf{k})]d k_y,
\end{aligned}
\end{equation} 
where $\mathcal{F}(\textbf{k})=\sum_{v=1,2}[| \varphi _{-v}(\textbf{k})\rangle \langle  \varphi _{-v}(\textbf{k})|-| \varphi _{v}(\textbf{k})\rangle \langle  \varphi _{v}(\textbf{k})|]$, with $| \varphi _{v}(\textbf{k})\rangle$ satisfying $\mathcal{H}(\textbf{k})| \varphi _{v}(\textbf{k})\rangle = E_{v}(\textbf{k})| \varphi _{v}(\textbf{k})\rangle$. Chiral symmetry guarantees $E_{-v}({\bf k})=-E_v({\bf k})$. Figure \ref{fig:2}(c) reveals that the regions with $W_{k_y}=\pm2$ correspond exactly to the regions enclosed by the boundary of the projection of the nodal lines in the $k_{x}$-$k_{z}$ plane, which form the drumhead surface states. Choosing $k_{z} = 0.76\pi$ in supporting the drumhead surface states, we plot in Fig. \ref{fig:2}(d) the dispersion relation under the boundary condition of open $y$ and closed $x$ directions. Matching well with Fig. \ref{fig:2}(c), the result exhibits two segments of flat bands signifying the drumhead surface states. The probability distribution of all the zero-mode states under the open boundary condition along both the $x$ and $y$ directions reveals that the surface and hinge Fermi arcs exist in the system [see Fig. \ref{fig:2}(e)], which verifies the second-order nature of the nodal lines. Thus, we realize an exotic second-order topological semimetal with coexisting nodal points and nodal lines.

\section{Floquet engineering}\label{fenggd}
The role of Floquet engineering lies in modifying the symmetry and inducing an effective long-range hopping within lattice systems \cite{PhysRevB.87.201109,PhysRevLett.121.036401}, which facilitates the exploration of novel topological phases completely absent in static systems. It reduces the experimental difficulty to generate topological phases via fabricating intrinsic interactions in natural materials. Hence, we introduce Floquet engineering into our system to enhance the diversity and controllability of our exotic topological semimetal. 

\begin{figure}[tbp]
\centering
\includegraphics[width=\columnwidth]{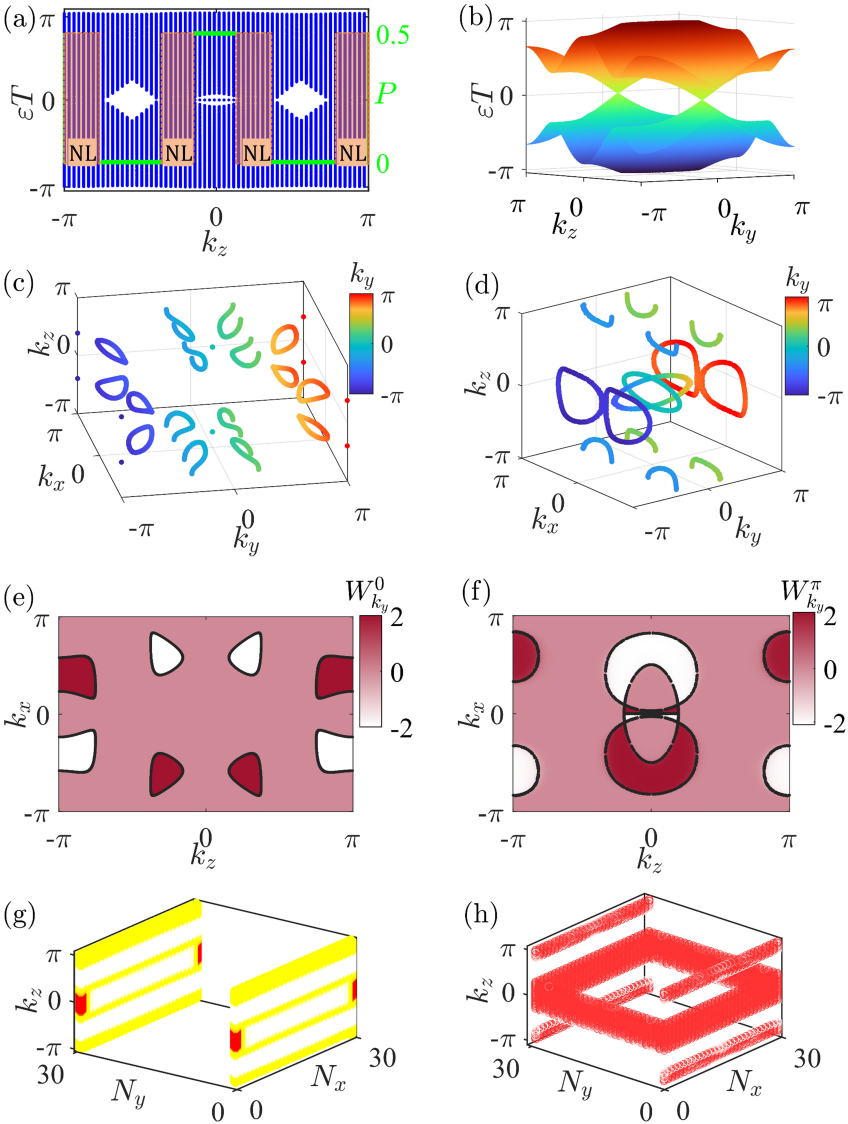}
\caption{(a) Quasienergy spectrum (blue dots) and quadrupole moment $P$ (green lines). The
nodal-lines areas marked by NL are shaded by orange. (b) Dispersion relations at $k_{x}=0$. Nodal-point and nodal-line distributions of the (c) zero mode and (d) $\pi/T$ mode and winding numbers (e) $W_{k_{y}}^{0}$ and (f) $W_{k_{y}}^{\pi}$ with the black lines denoting the projections of the nodal lines in the $k_{x}$-$k_{z}$ plane. Hybrid-order topological semimetal with coexisting second-order hinge Fermi arcs (red dots) and first-order surface Fermi arcs (yellow dots) in the zero mode (g) and purely first-order surface Fermi arcs in the $\pi/T$ mode (h). We use $\gamma=0.3f$, $\eta=0.6f$, $\lambda=0.1f$, $\mu_{1}=0.8$, $\mu_{2}=1.1$, $T_{1}=1.0f^{-1}$, and $T_{2}=1.5f^{-1}$.}\label{fig:3}
\end{figure}
We assume that $\mu$ varies periodically between $\mu_{1}$ and $\mu_{2}$ within the time durations $T_{1}$ and $T_{2}$, respectively, i.e.,
\begin{equation}\label{m0}
\begin{split}
\mu(t)= \left \{
 \begin{array}{ll}
\mu _{1}f,                    & t\in [lT,lT+T_{1})\\
\mu _{2}f,                    & t\in [lT+T_{1},(l+1)T),
 \end{array}
 \right.
\end{split}
\end{equation}
where $l\in\mathbb{Z}$, $T$=$T_{1}$+$T_{2}$ is the driving period, and $f$ is an energy scale to make the driving amplitudes $\mu_{1/2}$ dimensionless. Such a Floquet driving has also been employed to realize time crystals \cite{PhysRevLett.117.090402,PhysRevLett.118.030401}. Since the energy of the periodic system is not conserved, it does not have an energy spectrum. However, the Floquet theorem indicates that $|\phi_\alpha(t)\rangle$ and $\varepsilon_\alpha$ in the Floquet equation $[\hat{H}(t)-i\partial_t]|\phi_\alpha(t)\rangle=\varepsilon_\alpha|\phi_\alpha(t)\rangle$ are similar to the stationary states and eigenenergies of static systems, so they are called quasi-stationary states and quasienergies, respectively \cite{PhysRevA.7.2203,PhysRevA.91.052122}. The Floquet equation is equivalent to $\hat{U}_T|\phi_\alpha(0)\rangle=e^{-i\varepsilon_\alpha T}|\phi_\alpha(0)\rangle$, where $\hat{U}_T=\mathbb{T}e^{-i\int_0^T\hat{H}(t)dt}$ is a one-period evolution operator, with $\mathbb{T}$ being the time-ordering operator. Thus, $\hat{U}_T$ determines an effective static system $\hat{H}_\text{eff}=iT^{-1}\ln\hat{U}_T$, whose energy spectrum coincides with the quasienergy spectrum of our periodic system. Therefore, the tool for analyzing topological phases in static systems can be utilized to investigate periodic systems through $\hat{H}_\text{eff}$. Applying the Floquet theorem on our system, we have $\mathcal{H}_{\textrm{eff}}(\textbf{k})=iT^{-1}\,\textrm{ln}\,[e^{-i\mathcal{H}_{2}(\textbf{k})T_{2}}e^{-i\mathcal{H}_{1}(\textbf{k})T_{1}}]$, where $\mathcal{H}_{j}(\textbf{k})$ is the Hamiltonian \eqref{text} with $\mu$ replaced by $\mu_j$.

In contrast to the static case, the topological phases of periodic systems emerge at quasienergy gaps of not only zero, but also $\pi/T$, which makes the topological description well established in static systems insufficient for periodic systems. A complete topological description of our periodic system can be established as follows. The first-order topology is described by the winding number, which requires the presence of chiral symmetries. But $\mathcal{H}_{\textrm{eff}}(\textbf{k})$ does not inherit the chiral symmetry of the static system due to $[\mathcal{H}_{1}({\bf k}), \mathcal{H}_{2}({\bf k})]\neq 0$. We make two unitary transformations $D_{v}(\textbf{k})=e^{i(-1)^{v}\mathcal{H}_{v}(\textbf{k})T_{v}/2}$ ($v=1,2$), which do not change the quasienergy spectrum but recover the chiral symmetry, and obtain $\tilde{\mathcal{H}}_\text{eff,1}({\bf k})={iT^{-1}}\ln[e^{-i\mathcal{H}_1({\bf k})T_1/2}e^{-i\mathcal{H}_2({\bf k})T_2}e^{-i\mathcal{H}_1({\bf k})T_1/2}]$, $\tilde{\mathcal{H}}_\text{eff,2}({\bf k})={iT^{-1}}\ln[e^{-i\mathcal{H}_2({\bf k})T_2/2}e^{-i\mathcal{H}_1({\bf k})T_1}e^{-i\mathcal{H}_2({\bf k})T_2/2}]$ \cite{PhysRevB.103.L041115,PhysRevB.90.125143}. Then we can use the same method as the static system to define $W_{v,k_y}$ in $\tilde{\mathcal{H}}_\text{eff,v}({\bf k})$. The first-order topology of $\mathcal{H}_\text{eff}({\bf k})$ at the quasi-energies $\beta/T$, with $\beta=0$ or $\pi$, relate to $W_{v,k_y}$ as $W_{k_y}^{\beta}=(W_{1,k_y}+e^{i\beta } W_{2,k_y})/2$. Being similar to the static case, the second-order topology is characterized by the quadrupole moment.

The quasienergy spectrum of $\mathcal{H}_{\textrm{eff}}({\bf k})$ in Fig. \ref{fig:3}(a) reveals that the second-order corner states witnessed by $P=0.5$ are present at quasienergy zero but absent at $\pi/T$. The dispersion relation reveals the zero-mode Dirac points with four-fold degeneracy lying in $k_{x/y}=0$ [see Fig. \ref{fig:3}(b)]. The band touching points in Figs. \ref{fig:3}(c) and \ref{fig:3}(d) show the coexisting nodal points and lines at quasienergy zero but purely nodal lines in the quasienergy $\pi/T$. We see that the nodal-line structures are dramatically changed by periodic driving compared to the static case in Fig. \ref{fig:2}(b). A crossing-line node \cite{PhysRevB.95.245208}, where the two nodal loops in the planes of $k_x=0$ and $k_y=0$ are linked together, is formed in the $\pi/T$ mode. The winding numbers $W_{k_{y}}$ in Figs. \ref{fig:3}(e) and \ref{fig:3}(f) show that the system exhibits more fruitful first-order topological phases and drumhead surface states in the regions enclosed by the nodal loops in the quasienergy gap of the zero and $\pi/T$ modes than in the static case in Fig. \ref{fig:2}(c). Thus, the number of the drumhead surface states is dramatically enhanced. The probability distributions of the zero-mode states in Fig. \ref{fig:3}(g) indicate the second-order nature of the topological semimetal with second-order hinge Fermi arcs and first-order surface Fermi arcs. The ones of the $\pi/T$-mode states in Fig. \ref{fig:3}(h) reveal the first-order nature of the topological semimetal with pure surface Fermi arcs. Thus, we realize a hybrid-order topological semimetal that is first order with pure nodal lines at the $\pi/T$ mode and second order with coexisting second-order hinge and first-order surface Fermi arcs and coexisting nodal points and lines at the zero mode. Our hybrid-order topological semimetal has coexisting nodal point and lines, which is substantially different from the previously reported hybrid-order Weyl semimetal \cite{Wei_2024} and Floquet second-order Weyl semimetal \cite{PhysRevB.105.224312}. 

\begin{figure}[tbp]
\centering
\includegraphics[width=\columnwidth]{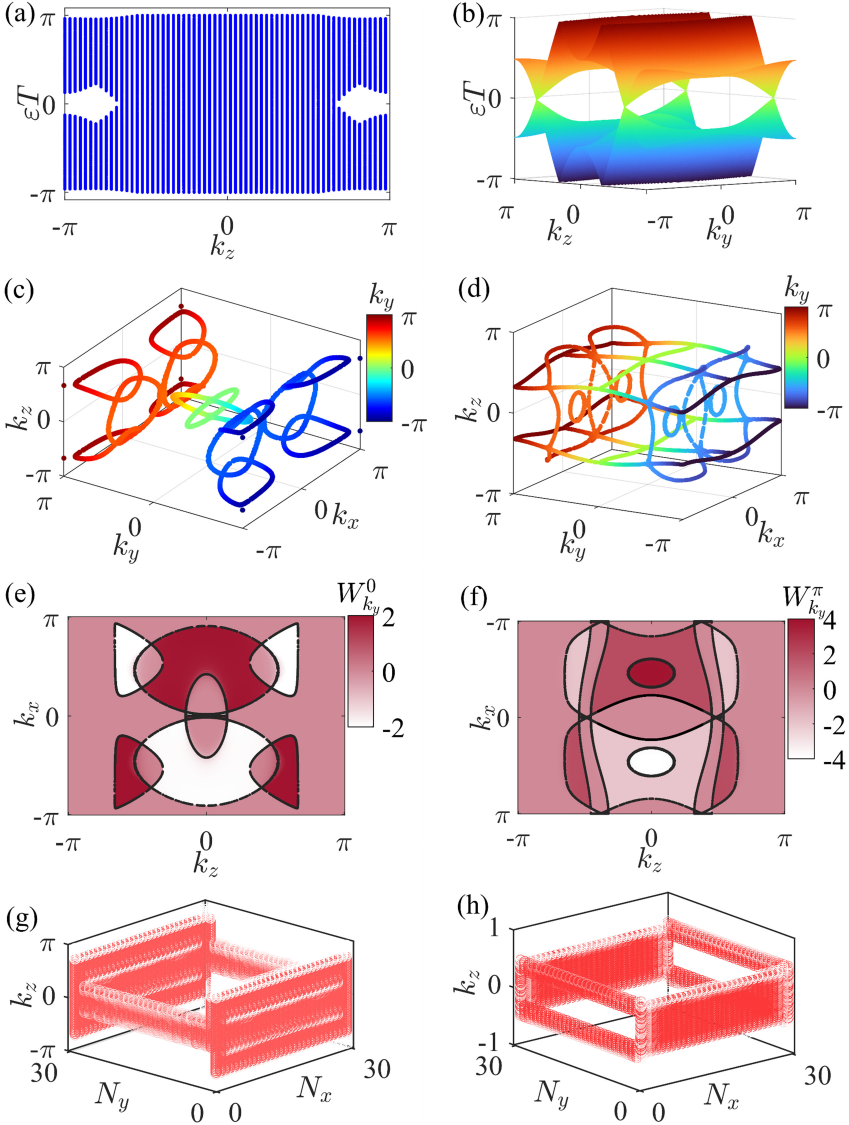}
\caption{(a) Quasienergy spectrum. (b) Dispersion relations at $k_{x}=\pi$. Nodal-point and nodal-line distributions of the (c) zero mode and (d) $\pi/T$ mode and winding numbers (e) $W_{k_{y}}^{0}$ and (f) $W_{k_{y}}^{\pi}$ with the black lines denoting the projections of the nodal lines on the $k_{x}$-$k_{z}$ plane. First-order topological semimetal with purely first-order surface Fermi arcs in the zero mode (g) and $\pi/T$ mode (h). We use $\gamma=0.2f$, $\lambda=0.5f$, $\eta=0.6f$, $\mu_{1}=0.85$, $\mu_{2}=1.9$, $T_{1}=2.4f^{-1}$, and $T_{2}=1.2f^{-1}$.}\label{fig:4}
\end{figure}

Moreover, the conversion between the different order topological semimetals can also be achieved through periodic driving, which is hard to realize in static systems. The quasienergy spectrum in Fig. \ref{fig:4}(a) shows that the zero-mode corner states disappear and no second-order phase is present in the 2D sliced system. The dispersion relation reveals that zero-mode Dirac points with four-fold degeneracy lie in $k_{x/y}=\pi$ [see Fig. \ref{fig:4}(b)]. Figures \ref{fig:4}(c) and \ref{fig:4}(d) show the distribution of the band touching of the zero and $\pi/T$ modes. We find that the structure of the nodal lines changes dramatically. Preserving the Dirac points at $k_{x/y}=\pm\pi$, the zero mode shows four nodal chains \cite{Bzdusek.2016} and a crossing-line node \cite{PhysRevB.95.245208}. The $\pi/T$ mode shows two nodal nets \cite{PhysRevLett.120.026402}, four separated nodal loops \cite{PhysRevLett.116.127202,PhysRevB.95.121107}, and two nodal chains \cite{Bzdusek.2016}. Figures \ref{fig:4}(e) and \ref{fig:4}(f) show the winding numbers $W_{k_y}^{0}$ and $W_{k_y}^{\pi}$. The topologically nontrivial regions match well with the ones enclosed by the nodal loops in Figs. \ref{fig:4}(c) and \ref{fig:4}(d). It is interesting to find that $W_{k_y}^\pi$ is widely changeable from $-4$ to $4$. It reveals that the numbers of drumhead surface states are enhanced. Both of the zero- and $\pi/T$-mode states are purely first-order surface Fermi arcs [see Figs. \ref{fig:3}(g) and \ref{fig:3}(h)]. The result indicates that the topological semimetal with coexisting nodal points and nodal lines undergoes a transition from second order in Fig. \ref{fig:3}(g) to first order. All the results prove that Floquet engineering offers us a useful tool to create and control novel topological semimetals by adding the time period as a new control dimension.

\section{Discussion and conclusion}\label{ddccl}
Note that the non-vanishing value of the interlayer hopping rate $\eta$ does not have qualitative influences on our results. Higher-order nodal-point semimetals have been realized in acoustic \cite{WeiQiang2021,Luo.Li.2021,PhysRevLett.127.146601,PhysRevLett.128.115701,PhysRevLett.130.116103}, photonic \cite{PhysRevB.105.L060101,PanYuang2023}, and electric-circuit \cite{SongLingling2022} systems. Higher-order nodal-line semimetals have been experimentally realized in acoustic systems \cite{PhysRevLett.132.197202,XueHaoran2023} and phononic crystals \cite{PhysRevLett.132.066601}. Dirac semimetals have been realized in Na$_{3}$Bi and Cd$_{3}$As$_{2}$ \cite{Z.K.Liu.2014,Neupane2014}. Weyl semimetals have been confirmed in TaAs \cite{Su-Yang.Xu.Ilya-Belopolski.2015}. Nodal-line semimetals have been observed in ZrSiS \cite{Schoop2016}. Floquet engineering has shown its ability to create exotic phases in photonic \cite{PhysRevLett.122.173901,Maczewsky.2017,Mukherjee.2017}, cold-atom \cite{Wintersperger.2010}, superconductor \cite{Roushan.P.2017}, and electronic-material \cite{Fahad.2016,McIver.2020} systems. Our model can be simulated by periodically modulating the acoustic coupling strength via an external piezoelectric element or acoustic actuation, modulating the refractive index of the photonic lattice, and modulating the coupling capacitance by microwave signals in electric-circuit systems \cite{10.1063/5.0150118,PhysRevLett.119.093901}. Our scheme is adaptable to other driving forms. The step-like driving protocol is used just for the convenience of numerical calculation, which avoids the difficulty in evaluating the time ordering of the evolution operator in a continuous harmonic driving \cite{PhysRevB.107.L121407}. These preexisting advances show that our predictions are realizable in currently available platforms and materials.

In summary, we have successfully devised a scheme to construct a unique second-order topological semimetal with coexisting nodal points and nodal lines and coexisting hinge Fermi arcs and drumhead surface states. In addition, through the application of periodic driving on our system, a hybrid-order topological semimetal with nodal points and rich nodal-line structures and its conversion to a purely first-order topological semimental can be realized on demand. Our work makes a substantial contribution to the promotion of the coexistence of different types and classes of topological semimetals. This expands the field and facilitates the development of topological semimetals. Furthermore, it provides a foundation for the design of multifunctional quantum devices through the controlled utilization of hybrid orders and classes of topological semimetals. 

\begin{acknowledgments}
The work is supported by the National Natural Science Foundation of China (Grants No. 124B2090, No. 12275109, and No. 12247101), the Innovation Program for Quantum Science and Technology of China (Grant No. 2023ZD0300904), the Fundamental Research Funds for the Central Universities (Grant No. lzujbky-2024-jdzx06), and the Natural Science Foundation of Gansu Province (No. 22JR5RA389).
\end{acknowledgments}

\textbf{Data availability.} The numerical data for generating the figures are available from authors upon request.

\bibliography{references}

\begin{thebibliography}{103}%
\makeatletter
\providecommand \@ifxundefined [1]{%
 \@ifx{#1\undefined}
}%
\providecommand \@ifnum [1]{%
 \ifnum #1\expandafter \@firstoftwo
 \else \expandafter \@secondoftwo
 \fi
}%
\providecommand \@ifx [1]{%
 \ifx #1\expandafter \@firstoftwo
 \else \expandafter \@secondoftwo
 \fi
}%
\providecommand \natexlab [1]{#1}%
\providecommand \enquote  [1]{``#1''}%
\providecommand \bibnamefont  [1]{#1}%
\providecommand \bibfnamefont [1]{#1}%
\providecommand \citenamefont [1]{#1}%
\providecommand \href@noop [0]{\@secondoftwo}%
\providecommand \href [0]{\begingroup \@sanitize@url \@href}%
\providecommand \@href[1]{\@@startlink{#1}\@@href}%
\providecommand \@@href[1]{\endgroup#1\@@endlink}%
\providecommand \@sanitize@url [0]{\catcode `\\12\catcode `\$12\catcode
  `\&12\catcode `\#12\catcode `\^12\catcode `\_12\catcode `\%12\relax}%
\providecommand \@@startlink[1]{}%
\providecommand \@@endlink[0]{}%
\providecommand \url  [0]{\begingroup\@sanitize@url \@url }%
\providecommand \@url [1]{\endgroup\@href {#1}{\urlprefix }}%
\providecommand \urlprefix  [0]{URL }%
\providecommand \Eprint [0]{\href }%
\providecommand \doibase [0]{https://doi.org/}%
\providecommand \selectlanguage [0]{\@gobble}%
\providecommand \bibinfo  [0]{\@secondoftwo}%
\providecommand \bibfield  [0]{\@secondoftwo}%
\providecommand \translation [1]{[#1]}%
\providecommand \BibitemOpen [0]{}%
\providecommand \bibitemStop [0]{}%
\providecommand \bibitemNoStop [0]{.\EOS\space}%
\providecommand \EOS [0]{\spacefactor3000\relax}%
\providecommand \BibitemShut  [1]{\csname bibitem#1\endcsname}%
\let\auto@bib@innerbib\@empty
\bibitem [{\citenamefont {Hasan}\ and\ \citenamefont
  {Kane}(2010)}]{RevModPhys.82.3045}%
  \BibitemOpen
  \bibfield  {author} {\bibinfo {author} {\bibfnamefont {M.~Z.}\ \bibnamefont
  {Hasan}}\ and\ \bibinfo {author} {\bibfnamefont {C.~L.}\ \bibnamefont
  {Kane}},\ }\bibfield  {title} {\bibinfo {title} {Colloquium: Topological
  insulators},\ }\href {https://doi.org/10.1103/RevModPhys.82.3045} {\bibfield
  {journal} {\bibinfo  {journal} {Rev. Mod. Phys.}\ }\textbf {\bibinfo {volume}
  {82}},\ \bibinfo {pages} {3045} (\bibinfo {year} {2010})}\BibitemShut
  {NoStop}%
\bibitem [{\citenamefont {Chiu}\ \emph {et~al.}(2016)\citenamefont {Chiu},
  \citenamefont {Teo}, \citenamefont {Schnyder},\ and\ \citenamefont
  {Ryu}}]{RevModPhys.88.035005}%
  \BibitemOpen
  \bibfield  {author} {\bibinfo {author} {\bibfnamefont {C.-K.}\ \bibnamefont
  {Chiu}}, \bibinfo {author} {\bibfnamefont {J.~C.~Y.}\ \bibnamefont {Teo}},
  \bibinfo {author} {\bibfnamefont {A.~P.}\ \bibnamefont {Schnyder}},\ and\
  \bibinfo {author} {\bibfnamefont {S.}~\bibnamefont {Ryu}},\ }\bibfield
  {title} {\bibinfo {title} {Classification of topological quantum matter with
  symmetries},\ }\href {https://doi.org/10.1103/RevModPhys.88.035005}
  {\bibfield  {journal} {\bibinfo  {journal} {Rev. Mod. Phys.}\ }\textbf
  {\bibinfo {volume} {88}},\ \bibinfo {pages} {035005} (\bibinfo {year}
  {2016})}\BibitemShut {NoStop}%
\bibitem [{\citenamefont {Qi}\ and\ \citenamefont
  {Zhang}(2011)}]{RevModPhys.83.1057}%
  \BibitemOpen
  \bibfield  {author} {\bibinfo {author} {\bibfnamefont {X.-L.}\ \bibnamefont
  {Qi}}\ and\ \bibinfo {author} {\bibfnamefont {S.-C.}\ \bibnamefont {Zhang}},\
  }\bibfield  {title} {\bibinfo {title} {Topological insulators and
  superconductors},\ }\href {https://doi.org/10.1103/RevModPhys.83.1057}
  {\bibfield  {journal} {\bibinfo  {journal} {Rev. Mod. Phys.}\ }\textbf
  {\bibinfo {volume} {83}},\ \bibinfo {pages} {1057} (\bibinfo {year}
  {2011})}\BibitemShut {NoStop}%
\bibitem [{\citenamefont {Armitage}\ \emph {et~al.}(2018)\citenamefont
  {Armitage}, \citenamefont {Mele},\ and\ \citenamefont
  {Vishwanath}}]{RevModPhys.90.015001}%
  \BibitemOpen
  \bibfield  {author} {\bibinfo {author} {\bibfnamefont {N.~P.}\ \bibnamefont
  {Armitage}}, \bibinfo {author} {\bibfnamefont {E.~J.}\ \bibnamefont {Mele}},\
  and\ \bibinfo {author} {\bibfnamefont {A.}~\bibnamefont {Vishwanath}},\
  }\bibfield  {title} {\bibinfo {title} {Weyl and {D}irac semimetals in
  three-dimensional solids},\ }\href
  {https://doi.org/10.1103/RevModPhys.90.015001} {\bibfield  {journal}
  {\bibinfo  {journal} {Rev. Mod. Phys.}\ }\textbf {\bibinfo {volume} {90}},\
  \bibinfo {pages} {015001} (\bibinfo {year} {2018})}\BibitemShut {NoStop}%
\bibitem [{\citenamefont {Lv}\ \emph {et~al.}(2021)\citenamefont {Lv},
  \citenamefont {Qian},\ and\ \citenamefont {Ding}}]{RevModPhys.93.025002}%
  \BibitemOpen
  \bibfield  {author} {\bibinfo {author} {\bibfnamefont {B.~Q.}\ \bibnamefont
  {Lv}}, \bibinfo {author} {\bibfnamefont {T.}~\bibnamefont {Qian}},\ and\
  \bibinfo {author} {\bibfnamefont {H.}~\bibnamefont {Ding}},\ }\bibfield
  {title} {\bibinfo {title} {Experimental perspective on three-dimensional
  topological semimetals},\ }\href
  {https://doi.org/10.1103/RevModPhys.93.025002} {\bibfield  {journal}
  {\bibinfo  {journal} {Rev. Mod. Phys.}\ }\textbf {\bibinfo {volume} {93}},\
  \bibinfo {pages} {025002} (\bibinfo {year} {2021})}\BibitemShut {NoStop}%
\bibitem [{\citenamefont {Benalcazar}\ \emph {et~al.}(2017)\citenamefont
  {Benalcazar}, \citenamefont {Bernevig},\ and\ \citenamefont
  {Hughes}}]{Benalcazar.2017}%
  \BibitemOpen
  \bibfield  {author} {\bibinfo {author} {\bibfnamefont {W.~A.}\ \bibnamefont
  {Benalcazar}}, \bibinfo {author} {\bibfnamefont {B.~A.}\ \bibnamefont
  {Bernevig}},\ and\ \bibinfo {author} {\bibfnamefont {T.~L.}\ \bibnamefont
  {Hughes}},\ }\bibfield  {title} {\bibinfo {title} {Quantized electric
  multipole insulators},\ }\href {https://doi.org/10.1126/science.aah6442}
  {\bibfield  {journal} {\bibinfo  {journal} {Science}\ }\textbf {\bibinfo
  {volume} {357}},\ \bibinfo {pages} {61} (\bibinfo {year} {2017})}\BibitemShut
  {NoStop}%
\bibitem [{\citenamefont {Xie}\ \emph {et~al.}(2021)\citenamefont {Xie},
  \citenamefont {Wang}, \citenamefont {Zhang}, \citenamefont {Zhan},
  \citenamefont {Jiang}, \citenamefont {Lu},\ and\ \citenamefont
  {Chen}}]{Xie.2021}%
  \BibitemOpen
  \bibfield  {author} {\bibinfo {author} {\bibfnamefont {B.}~\bibnamefont
  {Xie}}, \bibinfo {author} {\bibfnamefont {H.-X.}\ \bibnamefont {Wang}},
  \bibinfo {author} {\bibfnamefont {X.}~\bibnamefont {Zhang}}, \bibinfo
  {author} {\bibfnamefont {P.}~\bibnamefont {Zhan}}, \bibinfo {author}
  {\bibfnamefont {J.-H.}\ \bibnamefont {Jiang}}, \bibinfo {author}
  {\bibfnamefont {M.}~\bibnamefont {Lu}},\ and\ \bibinfo {author}
  {\bibfnamefont {Y.}~\bibnamefont {Chen}},\ }\bibfield  {title} {\bibinfo
  {title} {Higher-order band topology},\ }\href
  {https://doi.org/10.1038/s42254-021-00323-4} {\bibfield  {journal} {\bibinfo
  {journal} {Nature Reviews Physics}\ }\textbf {\bibinfo {volume} {3}},\
  \bibinfo {pages} {520} (\bibinfo {year} {2021})}\BibitemShut {NoStop}%
\bibitem [{\citenamefont {Langbehn}\ \emph {et~al.}(2017)\citenamefont
  {Langbehn}, \citenamefont {Peng}, \citenamefont {Trifunovic}, \citenamefont
  {von Oppen},\ and\ \citenamefont {Brouwer}}]{PhysRevLett.119.246401}%
  \BibitemOpen
  \bibfield  {author} {\bibinfo {author} {\bibfnamefont {J.}~\bibnamefont
  {Langbehn}}, \bibinfo {author} {\bibfnamefont {Y.}~\bibnamefont {Peng}},
  \bibinfo {author} {\bibfnamefont {L.}~\bibnamefont {Trifunovic}}, \bibinfo
  {author} {\bibfnamefont {F.}~\bibnamefont {von Oppen}},\ and\ \bibinfo
  {author} {\bibfnamefont {P.~W.}\ \bibnamefont {Brouwer}},\ }\bibfield
  {title} {\bibinfo {title} {Reflection-{S}ymmetric {S}econd-order
  {T}opological {I}nsulators and {S}uperconductors},\ }\href
  {https://doi.org/10.1103/PhysRevLett.119.246401} {\bibfield  {journal}
  {\bibinfo  {journal} {Phys. Rev. Lett.}\ }\textbf {\bibinfo {volume} {119}},\
  \bibinfo {pages} {246401} (\bibinfo {year} {2017})}\BibitemShut {NoStop}%
\bibitem [{\citenamefont {Fan}\ \emph {et~al.}(2019)\citenamefont {Fan},
  \citenamefont {Xia}, \citenamefont {Tong}, \citenamefont {Zheng},\ and\
  \citenamefont {Yu}}]{PhysRevLett.122.204301}%
  \BibitemOpen
  \bibfield  {author} {\bibinfo {author} {\bibfnamefont {H.}~\bibnamefont
  {Fan}}, \bibinfo {author} {\bibfnamefont {B.}~\bibnamefont {Xia}}, \bibinfo
  {author} {\bibfnamefont {L.}~\bibnamefont {Tong}}, \bibinfo {author}
  {\bibfnamefont {S.}~\bibnamefont {Zheng}},\ and\ \bibinfo {author}
  {\bibfnamefont {D.}~\bibnamefont {Yu}},\ }\bibfield  {title} {\bibinfo
  {title} {Elastic {H}igher-order {T}opological {I}nsulator with
  {T}opologically {P}rotected {C}orner {S}tates},\ }\href
  {https://doi.org/10.1103/PhysRevLett.122.204301} {\bibfield  {journal}
  {\bibinfo  {journal} {Phys. Rev. Lett.}\ }\textbf {\bibinfo {volume} {122}},\
  \bibinfo {pages} {204301} (\bibinfo {year} {2019})}\BibitemShut {NoStop}%
\bibitem [{\citenamefont {Ghosh}\ \emph {et~al.}(2023)\citenamefont {Ghosh},
  \citenamefont {Nag},\ and\ \citenamefont {Saha}}]{Ghosh_2024}%
  \BibitemOpen
  \bibfield  {author} {\bibinfo {author} {\bibfnamefont {A.~K.}\ \bibnamefont
  {Ghosh}}, \bibinfo {author} {\bibfnamefont {T.}~\bibnamefont {Nag}},\ and\
  \bibinfo {author} {\bibfnamefont {A.}~\bibnamefont {Saha}},\ }\bibfield
  {title} {\bibinfo {title} {Generation of higher-order topological insulators
  using periodic driving},\ }\href {https://doi.org/10.1088/1361-648X/ad0e2d}
  {\bibfield  {journal} {\bibinfo  {journal} {Journal of Physics: Condensed
  Matter}\ }\textbf {\bibinfo {volume} {36}},\ \bibinfo {pages} {093001}
  (\bibinfo {year} {2023})}\BibitemShut {NoStop}%
\bibitem [{\citenamefont {Young}\ \emph {et~al.}(2012)\citenamefont {Young},
  \citenamefont {Zaheer}, \citenamefont {Teo}, \citenamefont {Kane},
  \citenamefont {Mele},\ and\ \citenamefont {Rappe}}]{PhysRevLett.108.140405}%
  \BibitemOpen
  \bibfield  {author} {\bibinfo {author} {\bibfnamefont {S.~M.}\ \bibnamefont
  {Young}}, \bibinfo {author} {\bibfnamefont {S.}~\bibnamefont {Zaheer}},
  \bibinfo {author} {\bibfnamefont {J.~C.~Y.}\ \bibnamefont {Teo}}, \bibinfo
  {author} {\bibfnamefont {C.~L.}\ \bibnamefont {Kane}}, \bibinfo {author}
  {\bibfnamefont {E.~J.}\ \bibnamefont {Mele}},\ and\ \bibinfo {author}
  {\bibfnamefont {A.~M.}\ \bibnamefont {Rappe}},\ }\bibfield  {title} {\bibinfo
  {title} {Dirac {S}emimetal in {T}hree {D}imensions},\ }\href
  {https://doi.org/10.1103/PhysRevLett.108.140405} {\bibfield  {journal}
  {\bibinfo  {journal} {Phys. Rev. Lett.}\ }\textbf {\bibinfo {volume} {108}},\
  \bibinfo {pages} {140405} (\bibinfo {year} {2012})}\BibitemShut {NoStop}%
\bibitem [{\citenamefont {Young}\ and\ \citenamefont
  {Kane}(2015)}]{PhysRevLett.115.126803}%
  \BibitemOpen
  \bibfield  {author} {\bibinfo {author} {\bibfnamefont {S.~M.}\ \bibnamefont
  {Young}}\ and\ \bibinfo {author} {\bibfnamefont {C.~L.}\ \bibnamefont
  {Kane}},\ }\bibfield  {title} {\bibinfo {title} {Dirac {S}emimetals in {T}wo
  {D}imensions},\ }\href {https://doi.org/10.1103/PhysRevLett.115.126803}
  {\bibfield  {journal} {\bibinfo  {journal} {Phys. Rev. Lett.}\ }\textbf
  {\bibinfo {volume} {115}},\ \bibinfo {pages} {126803} (\bibinfo {year}
  {2015})}\BibitemShut {NoStop}%
\bibitem [{\citenamefont {Chang}\ \emph {et~al.}(2017)\citenamefont {Chang},
  \citenamefont {Xu}, \citenamefont {Sanchez}, \citenamefont {Tsai},
  \citenamefont {Huang}, \citenamefont {Chang}, \citenamefont {Hsu},
  \citenamefont {Bian}, \citenamefont {Belopolski}, \citenamefont {Yu},
  \citenamefont {Yang}, \citenamefont {Neupert}, \citenamefont {Jeng},
  \citenamefont {Lin},\ and\ \citenamefont {Hasan}}]{PhysRevLett.119.026404}%
  \BibitemOpen
  \bibfield  {author} {\bibinfo {author} {\bibfnamefont {T.-R.}\ \bibnamefont
  {Chang}}, \bibinfo {author} {\bibfnamefont {S.-Y.}\ \bibnamefont {Xu}},
  \bibinfo {author} {\bibfnamefont {D.~S.}\ \bibnamefont {Sanchez}}, \bibinfo
  {author} {\bibfnamefont {W.-F.}\ \bibnamefont {Tsai}}, \bibinfo {author}
  {\bibfnamefont {S.-M.}\ \bibnamefont {Huang}}, \bibinfo {author}
  {\bibfnamefont {G.}~\bibnamefont {Chang}}, \bibinfo {author} {\bibfnamefont
  {C.-H.}\ \bibnamefont {Hsu}}, \bibinfo {author} {\bibfnamefont
  {G.}~\bibnamefont {Bian}}, \bibinfo {author} {\bibfnamefont {I.}~\bibnamefont
  {Belopolski}}, \bibinfo {author} {\bibfnamefont {Z.-M.}\ \bibnamefont {Yu}},
  \bibinfo {author} {\bibfnamefont {S.~A.}\ \bibnamefont {Yang}}, \bibinfo
  {author} {\bibfnamefont {T.}~\bibnamefont {Neupert}}, \bibinfo {author}
  {\bibfnamefont {H.-T.}\ \bibnamefont {Jeng}}, \bibinfo {author}
  {\bibfnamefont {H.}~\bibnamefont {Lin}},\ and\ \bibinfo {author}
  {\bibfnamefont {M.~Z.}\ \bibnamefont {Hasan}},\ }\bibfield  {title} {\bibinfo
  {title} {Type-{II} {S}ymmetry-{P}rotected {T}opological {D}irac
  {S}emimetals},\ }\href {https://doi.org/10.1103/PhysRevLett.119.026404}
  {\bibfield  {journal} {\bibinfo  {journal} {Phys. Rev. Lett.}\ }\textbf
  {\bibinfo {volume} {119}},\ \bibinfo {pages} {026404} (\bibinfo {year}
  {2017})}\BibitemShut {NoStop}%
\bibitem [{\citenamefont {Liu}\ \emph {et~al.}(2014)\citenamefont {Liu},
  \citenamefont {Zhou}, \citenamefont {Zhang}, \citenamefont {Wang},
  \citenamefont {Weng}, \citenamefont {Prabhakaran}, \citenamefont {Mo},
  \citenamefont {Shen}, \citenamefont {Fang}, \citenamefont {Dai},
  \citenamefont {Hussain},\ and\ \citenamefont {Chen}}]{Z.K.Liu.2014}%
  \BibitemOpen
  \bibfield  {author} {\bibinfo {author} {\bibfnamefont {Z.~K.}\ \bibnamefont
  {Liu}}, \bibinfo {author} {\bibfnamefont {B.}~\bibnamefont {Zhou}}, \bibinfo
  {author} {\bibfnamefont {Y.}~\bibnamefont {Zhang}}, \bibinfo {author}
  {\bibfnamefont {Z.~J.}\ \bibnamefont {Wang}}, \bibinfo {author}
  {\bibfnamefont {H.~M.}\ \bibnamefont {Weng}}, \bibinfo {author}
  {\bibfnamefont {D.}~\bibnamefont {Prabhakaran}}, \bibinfo {author}
  {\bibfnamefont {S.-K.}\ \bibnamefont {Mo}}, \bibinfo {author} {\bibfnamefont
  {Z.~X.}\ \bibnamefont {Shen}}, \bibinfo {author} {\bibfnamefont
  {Z.}~\bibnamefont {Fang}}, \bibinfo {author} {\bibfnamefont {X.}~\bibnamefont
  {Dai}}, \bibinfo {author} {\bibfnamefont {Z.}~\bibnamefont {Hussain}},\ and\
  \bibinfo {author} {\bibfnamefont {Y.~L.}\ \bibnamefont {Chen}},\ }\bibfield
  {title} {\bibinfo {title} {Discovery of a {T}hree-{D}imensional {T}opological
  {D}irac {S}emimetal, {N}$\textrm{a}_{3}${B}i},\ }\href
  {https://doi.org/10.1126/science.1245085} {\bibfield  {journal} {\bibinfo
  {journal} {Science}\ }\textbf {\bibinfo {volume} {343}},\ \bibinfo {pages}
  {864} (\bibinfo {year} {2014})}\BibitemShut {NoStop}%
\bibitem [{\citenamefont {Borisenko}\ \emph {et~al.}(2014)\citenamefont
  {Borisenko}, \citenamefont {Gibson}, \citenamefont {Evtushinsky},
  \citenamefont {Zabolotnyy}, \citenamefont {B\"uchner},\ and\ \citenamefont
  {Cava}}]{PhysRevLett.113.027603}%
  \BibitemOpen
  \bibfield  {author} {\bibinfo {author} {\bibfnamefont {S.}~\bibnamefont
  {Borisenko}}, \bibinfo {author} {\bibfnamefont {Q.}~\bibnamefont {Gibson}},
  \bibinfo {author} {\bibfnamefont {D.}~\bibnamefont {Evtushinsky}}, \bibinfo
  {author} {\bibfnamefont {V.}~\bibnamefont {Zabolotnyy}}, \bibinfo {author}
  {\bibfnamefont {B.}~\bibnamefont {B\"uchner}},\ and\ \bibinfo {author}
  {\bibfnamefont {R.~J.}\ \bibnamefont {Cava}},\ }\bibfield  {title} {\bibinfo
  {title} {Experimental {R}ealization of a {T}hree-{D}imensional {D}irac
  {S}emimetal},\ }\href {https://doi.org/10.1103/PhysRevLett.113.027603}
  {\bibfield  {journal} {\bibinfo  {journal} {Phys. Rev. Lett.}\ }\textbf
  {\bibinfo {volume} {113}},\ \bibinfo {pages} {027603} (\bibinfo {year}
  {2014})}\BibitemShut {NoStop}%
\bibitem [{\citenamefont {Wieder}\ \emph {et~al.}(2020)\citenamefont {Wieder},
  \citenamefont {Wang}, \citenamefont {Cano}, \citenamefont {Dai},
  \citenamefont {Schoop}, \citenamefont {Bradlyn},\ and\ \citenamefont
  {Bernevig}}]{Wieder.Benjamin.J.2020}%
  \BibitemOpen
  \bibfield  {author} {\bibinfo {author} {\bibfnamefont {B.~J.}\ \bibnamefont
  {Wieder}}, \bibinfo {author} {\bibfnamefont {Z.}~\bibnamefont {Wang}},
  \bibinfo {author} {\bibfnamefont {J.}~\bibnamefont {Cano}}, \bibinfo {author}
  {\bibfnamefont {X.}~\bibnamefont {Dai}}, \bibinfo {author} {\bibfnamefont
  {L.~M.}\ \bibnamefont {Schoop}}, \bibinfo {author} {\bibfnamefont
  {B.}~\bibnamefont {Bradlyn}},\ and\ \bibinfo {author} {\bibfnamefont {B.~A.}\
  \bibnamefont {Bernevig}},\ }\bibfield  {title} {\bibinfo {title} {Strong and
  fragile topological {D}irac {s}emimetals with higher-order {F}ermi arcs},\
  }\href {https://doi.org/10.1038/s41467-020-14443-5} {\bibfield  {journal}
  {\bibinfo  {journal} {Nature Communications}\ }\textbf {\bibinfo {volume}
  {11}},\ \bibinfo {pages} {627} (\bibinfo {year} {2020})}\BibitemShut
  {NoStop}%
\bibitem [{\citenamefont {Xu}\ \emph {et~al.}(2015)\citenamefont {Xu},
  \citenamefont {Belopolski}, \citenamefont {Alidoust}, \citenamefont
  {Neupane}, \citenamefont {Bian}, \citenamefont {Zhang}, \citenamefont
  {Sankar}, \citenamefont {Chang}, \citenamefont {Yuan}, \citenamefont {Lee},
  \citenamefont {Huang}, \citenamefont {Zheng}, \citenamefont {Ma},
  \citenamefont {Sanchez}, \citenamefont {Wang}, \citenamefont {Bansil},
  \citenamefont {Chou}, \citenamefont {Shibayev}, \citenamefont {Lin},
  \citenamefont {Jia},\ and\ \citenamefont
  {Hasan}}]{Su-Yang.Xu.Ilya-Belopolski.2015}%
  \BibitemOpen
  \bibfield  {author} {\bibinfo {author} {\bibfnamefont {S.-Y.}\ \bibnamefont
  {Xu}}, \bibinfo {author} {\bibfnamefont {I.}~\bibnamefont {Belopolski}},
  \bibinfo {author} {\bibfnamefont {N.}~\bibnamefont {Alidoust}}, \bibinfo
  {author} {\bibfnamefont {M.}~\bibnamefont {Neupane}}, \bibinfo {author}
  {\bibfnamefont {G.}~\bibnamefont {Bian}}, \bibinfo {author} {\bibfnamefont
  {C.}~\bibnamefont {Zhang}}, \bibinfo {author} {\bibfnamefont
  {R.}~\bibnamefont {Sankar}}, \bibinfo {author} {\bibfnamefont
  {G.}~\bibnamefont {Chang}}, \bibinfo {author} {\bibfnamefont
  {Z.}~\bibnamefont {Yuan}}, \bibinfo {author} {\bibfnamefont {C.-C.}\
  \bibnamefont {Lee}}, \bibinfo {author} {\bibfnamefont {S.-M.}\ \bibnamefont
  {Huang}}, \bibinfo {author} {\bibfnamefont {H.}~\bibnamefont {Zheng}},
  \bibinfo {author} {\bibfnamefont {J.}~\bibnamefont {Ma}}, \bibinfo {author}
  {\bibfnamefont {D.~S.}\ \bibnamefont {Sanchez}}, \bibinfo {author}
  {\bibfnamefont {B.}~\bibnamefont {Wang}}, \bibinfo {author} {\bibfnamefont
  {A.}~\bibnamefont {Bansil}}, \bibinfo {author} {\bibfnamefont
  {F.}~\bibnamefont {Chou}}, \bibinfo {author} {\bibfnamefont {P.~P.}\
  \bibnamefont {Shibayev}}, \bibinfo {author} {\bibfnamefont {H.}~\bibnamefont
  {Lin}}, \bibinfo {author} {\bibfnamefont {S.}~\bibnamefont {Jia}},\ and\
  \bibinfo {author} {\bibfnamefont {M.~Z.}\ \bibnamefont {Hasan}},\ }\bibfield
  {title} {\bibinfo {title} {Discovery of a {W}eyl fermion semimetal and
  topological {F}ermi arcs},\ }\href {https://doi.org/10.1126/science.aaa9297}
  {\bibfield  {journal} {\bibinfo  {journal} {Science}\ }\textbf {\bibinfo
  {volume} {349}},\ \bibinfo {pages} {613} (\bibinfo {year}
  {2015})}\BibitemShut {NoStop}%
\bibitem [{\citenamefont {Lv}\ \emph {et~al.}(2015)\citenamefont {Lv},
  \citenamefont {Weng}, \citenamefont {Fu}, \citenamefont {Wang}, \citenamefont
  {Miao}, \citenamefont {Ma}, \citenamefont {Richard}, \citenamefont {Huang},
  \citenamefont {Zhao}, \citenamefont {Chen}, \citenamefont {Fang},
  \citenamefont {Dai}, \citenamefont {Qian},\ and\ \citenamefont
  {Ding}}]{PhysRevX.5.031013}%
  \BibitemOpen
  \bibfield  {author} {\bibinfo {author} {\bibfnamefont {B.~Q.}\ \bibnamefont
  {Lv}}, \bibinfo {author} {\bibfnamefont {H.~M.}\ \bibnamefont {Weng}},
  \bibinfo {author} {\bibfnamefont {B.~B.}\ \bibnamefont {Fu}}, \bibinfo
  {author} {\bibfnamefont {X.~P.}\ \bibnamefont {Wang}}, \bibinfo {author}
  {\bibfnamefont {H.}~\bibnamefont {Miao}}, \bibinfo {author} {\bibfnamefont
  {J.}~\bibnamefont {Ma}}, \bibinfo {author} {\bibfnamefont {P.}~\bibnamefont
  {Richard}}, \bibinfo {author} {\bibfnamefont {X.~C.}\ \bibnamefont {Huang}},
  \bibinfo {author} {\bibfnamefont {L.~X.}\ \bibnamefont {Zhao}}, \bibinfo
  {author} {\bibfnamefont {G.~F.}\ \bibnamefont {Chen}}, \bibinfo {author}
  {\bibfnamefont {Z.}~\bibnamefont {Fang}}, \bibinfo {author} {\bibfnamefont
  {X.}~\bibnamefont {Dai}}, \bibinfo {author} {\bibfnamefont {T.}~\bibnamefont
  {Qian}},\ and\ \bibinfo {author} {\bibfnamefont {H.}~\bibnamefont {Ding}},\
  }\bibfield  {title} {\bibinfo {title} {Experimental {D}iscovery of {W}eyl
  {S}emimetal {T}a{A}s},\ }\href {https://doi.org/10.1103/PhysRevX.5.031013}
  {\bibfield  {journal} {\bibinfo  {journal} {Phys. Rev. X}\ }\textbf {\bibinfo
  {volume} {5}},\ \bibinfo {pages} {031013} (\bibinfo {year}
  {2015})}\BibitemShut {NoStop}%
\bibitem [{\citenamefont {Yang}\ \emph {et~al.}(2015)\citenamefont {Yang},
  \citenamefont {Liu}, \citenamefont {Sun}, \citenamefont {Peng}, \citenamefont
  {Yang}, \citenamefont {Zhang}, \citenamefont {Zhou}, \citenamefont {Zhang},
  \citenamefont {Guo}, \citenamefont {Rahn}, \citenamefont {Prabhakaran},
  \citenamefont {Hussain}, \citenamefont {Mo}, \citenamefont {Felser},
  \citenamefont {Yan},\ and\ \citenamefont {Chen}}]{Yang.L.X.2015}%
  \BibitemOpen
  \bibfield  {author} {\bibinfo {author} {\bibfnamefont {L.~X.}\ \bibnamefont
  {Yang}}, \bibinfo {author} {\bibfnamefont {Z.~K.}\ \bibnamefont {Liu}},
  \bibinfo {author} {\bibfnamefont {Y.}~\bibnamefont {Sun}}, \bibinfo {author}
  {\bibfnamefont {H.}~\bibnamefont {Peng}}, \bibinfo {author} {\bibfnamefont
  {H.~F.}\ \bibnamefont {Yang}}, \bibinfo {author} {\bibfnamefont
  {T.}~\bibnamefont {Zhang}}, \bibinfo {author} {\bibfnamefont
  {B.}~\bibnamefont {Zhou}}, \bibinfo {author} {\bibfnamefont {Y.}~\bibnamefont
  {Zhang}}, \bibinfo {author} {\bibfnamefont {Y.~F.}\ \bibnamefont {Guo}},
  \bibinfo {author} {\bibfnamefont {M.}~\bibnamefont {Rahn}}, \bibinfo {author}
  {\bibfnamefont {D.}~\bibnamefont {Prabhakaran}}, \bibinfo {author}
  {\bibfnamefont {Z.}~\bibnamefont {Hussain}}, \bibinfo {author} {\bibfnamefont
  {S.~K.}\ \bibnamefont {Mo}}, \bibinfo {author} {\bibfnamefont
  {C.}~\bibnamefont {Felser}}, \bibinfo {author} {\bibfnamefont
  {B.}~\bibnamefont {Yan}},\ and\ \bibinfo {author} {\bibfnamefont {Y.~L.}\
  \bibnamefont {Chen}},\ }\bibfield  {title} {\bibinfo {title} {Weyl semimetal
  phase in the non-centrosymmetric compound {T}a{A}s},\ }\href
  {https://doi.org/10.1038/nphys3425} {\bibfield  {journal} {\bibinfo
  {journal} {Nature Physics}\ }\textbf {\bibinfo {volume} {11}},\ \bibinfo
  {pages} {728} (\bibinfo {year} {2015})}\BibitemShut {NoStop}%
\bibitem [{\citenamefont {Wang}\ \emph
  {et~al.}(2020{\natexlab{a}})\citenamefont {Wang}, \citenamefont {Lin},
  \citenamefont {Jiang}, \citenamefont {Guo},\ and\ \citenamefont
  {Jiang}}]{PhysRevLett.125.146401}%
  \BibitemOpen
  \bibfield  {author} {\bibinfo {author} {\bibfnamefont {H.-X.}\ \bibnamefont
  {Wang}}, \bibinfo {author} {\bibfnamefont {Z.-K.}\ \bibnamefont {Lin}},
  \bibinfo {author} {\bibfnamefont {B.}~\bibnamefont {Jiang}}, \bibinfo
  {author} {\bibfnamefont {G.-Y.}\ \bibnamefont {Guo}},\ and\ \bibinfo {author}
  {\bibfnamefont {J.-H.}\ \bibnamefont {Jiang}},\ }\bibfield  {title} {\bibinfo
  {title} {Higher-{O}rder {W}eyl {S}emimetals},\ }\href
  {https://doi.org/10.1103/PhysRevLett.125.146401} {\bibfield  {journal}
  {\bibinfo  {journal} {Phys. Rev. Lett.}\ }\textbf {\bibinfo {volume} {125}},\
  \bibinfo {pages} {146401} (\bibinfo {year} {2020}{\natexlab{a}})}\BibitemShut
  {NoStop}%
\bibitem [{\citenamefont {Ghorashi}\ \emph {et~al.}(2020)\citenamefont
  {Ghorashi}, \citenamefont {Li},\ and\ \citenamefont
  {Hughes}}]{PhysRevLett.125.266804}%
  \BibitemOpen
  \bibfield  {author} {\bibinfo {author} {\bibfnamefont {S.~A.~A.}\
  \bibnamefont {Ghorashi}}, \bibinfo {author} {\bibfnamefont {T.}~\bibnamefont
  {Li}},\ and\ \bibinfo {author} {\bibfnamefont {T.~L.}\ \bibnamefont
  {Hughes}},\ }\bibfield  {title} {\bibinfo {title} {Higher-{O}rder {W}eyl
  {S}emimetals},\ }\href {https://doi.org/10.1103/PhysRevLett.125.266804}
  {\bibfield  {journal} {\bibinfo  {journal} {Phys. Rev. Lett.}\ }\textbf
  {\bibinfo {volume} {125}},\ \bibinfo {pages} {266804} (\bibinfo {year}
  {2020})}\BibitemShut {NoStop}%
\bibitem [{\citenamefont {Luo}\ \emph {et~al.}(2021)\citenamefont {Luo},
  \citenamefont {Wang}, \citenamefont {Lin}, \citenamefont {Jiang},
  \citenamefont {Wu}, \citenamefont {Li},\ and\ \citenamefont
  {Jiang}}]{Luo.Li.2021}%
  \BibitemOpen
  \bibfield  {author} {\bibinfo {author} {\bibfnamefont {L.}~\bibnamefont
  {Luo}}, \bibinfo {author} {\bibfnamefont {H.-X.}\ \bibnamefont {Wang}},
  \bibinfo {author} {\bibfnamefont {Z.-K.}\ \bibnamefont {Lin}}, \bibinfo
  {author} {\bibfnamefont {B.}~\bibnamefont {Jiang}}, \bibinfo {author}
  {\bibfnamefont {Y.}~\bibnamefont {Wu}}, \bibinfo {author} {\bibfnamefont
  {F.}~\bibnamefont {Li}},\ and\ \bibinfo {author} {\bibfnamefont {J.-H.}\
  \bibnamefont {Jiang}},\ }\bibfield  {title} {\bibinfo {title} {Observation of
  a phononic higher-order {W}eyl semimetal},\ }\href
  {https://doi.org/10.1038/s41563-021-00985-6} {\bibfield  {journal} {\bibinfo
  {journal} {Nature Materials}\ }\textbf {\bibinfo {volume} {20}},\ \bibinfo
  {pages} {794} (\bibinfo {year} {2021})}\BibitemShut {NoStop}%
\bibitem [{\citenamefont {Xiong}\ \emph {et~al.}(2024)\citenamefont {Xiong},
  \citenamefont {He}, \citenamefont {Liu}, \citenamefont {Black-Schaffer},\
  and\ \citenamefont {Nag}}]{PhysRevB.109.054201}%
  \BibitemOpen
  \bibfield  {author} {\bibinfo {author} {\bibfnamefont {F.}~\bibnamefont
  {Xiong}}, \bibinfo {author} {\bibfnamefont {C.}~\bibnamefont {He}}, \bibinfo
  {author} {\bibfnamefont {Y.}~\bibnamefont {Liu}}, \bibinfo {author}
  {\bibfnamefont {A.~M.}\ \bibnamefont {Black-Schaffer}},\ and\ \bibinfo
  {author} {\bibfnamefont {T.}~\bibnamefont {Nag}},\ }\bibfield  {title}
  {\bibinfo {title} {Distinct quasiparticle interference patterns for surface
  impurity scattering on various weyl semimetals},\ }\href
  {https://doi.org/10.1103/PhysRevB.109.054201} {\bibfield  {journal} {\bibinfo
   {journal} {Phys. Rev. B}\ }\textbf {\bibinfo {volume} {109}},\ \bibinfo
  {pages} {054201} (\bibinfo {year} {2024})}\BibitemShut {NoStop}%
\bibitem [{\citenamefont {Xiong}\ \emph {et~al.}(2022)\citenamefont {Xiong},
  \citenamefont {Honerkamp}, \citenamefont {Kennes},\ and\ \citenamefont
  {Nag}}]{PhysRevB.106.045424}%
  \BibitemOpen
  \bibfield  {author} {\bibinfo {author} {\bibfnamefont {F.}~\bibnamefont
  {Xiong}}, \bibinfo {author} {\bibfnamefont {C.}~\bibnamefont {Honerkamp}},
  \bibinfo {author} {\bibfnamefont {D.~M.}\ \bibnamefont {Kennes}},\ and\
  \bibinfo {author} {\bibfnamefont {T.}~\bibnamefont {Nag}},\ }\bibfield
  {title} {\bibinfo {title} {Understanding the three-dimensional quantum hall
  effect in generic multi-weyl semimetals},\ }\href
  {https://doi.org/10.1103/PhysRevB.106.045424} {\bibfield  {journal} {\bibinfo
   {journal} {Phys. Rev. B}\ }\textbf {\bibinfo {volume} {106}},\ \bibinfo
  {pages} {045424} (\bibinfo {year} {2022})}\BibitemShut {NoStop}%
\bibitem [{\citenamefont {Burkov}\ \emph {et~al.}(2011)\citenamefont {Burkov},
  \citenamefont {Hook},\ and\ \citenamefont {Balents}}]{PhysRevB.84.235126}%
  \BibitemOpen
  \bibfield  {author} {\bibinfo {author} {\bibfnamefont {A.~A.}\ \bibnamefont
  {Burkov}}, \bibinfo {author} {\bibfnamefont {M.~D.}\ \bibnamefont {Hook}},\
  and\ \bibinfo {author} {\bibfnamefont {L.}~\bibnamefont {Balents}},\
  }\bibfield  {title} {\bibinfo {title} {Topological nodal semimetals},\ }\href
  {https://doi.org/10.1103/PhysRevB.84.235126} {\bibfield  {journal} {\bibinfo
  {journal} {Phys. Rev. B}\ }\textbf {\bibinfo {volume} {84}},\ \bibinfo
  {pages} {235126} (\bibinfo {year} {2011})}\BibitemShut {NoStop}%
\bibitem [{\citenamefont {Yan}\ \emph {et~al.}(2017)\citenamefont {Yan},
  \citenamefont {Bi}, \citenamefont {Shen}, \citenamefont {Lu}, \citenamefont
  {Zhang},\ and\ \citenamefont {Wang}}]{PhysRevB.96.041103}%
  \BibitemOpen
  \bibfield  {author} {\bibinfo {author} {\bibfnamefont {Z.}~\bibnamefont
  {Yan}}, \bibinfo {author} {\bibfnamefont {R.}~\bibnamefont {Bi}}, \bibinfo
  {author} {\bibfnamefont {H.}~\bibnamefont {Shen}}, \bibinfo {author}
  {\bibfnamefont {L.}~\bibnamefont {Lu}}, \bibinfo {author} {\bibfnamefont
  {S.-C.}\ \bibnamefont {Zhang}},\ and\ \bibinfo {author} {\bibfnamefont
  {Z.}~\bibnamefont {Wang}},\ }\bibfield  {title} {\bibinfo {title} {Nodal-link
  semimetals},\ }\href {https://doi.org/10.1103/PhysRevB.96.041103} {\bibfield
  {journal} {\bibinfo  {journal} {Phys. Rev. B}\ }\textbf {\bibinfo {volume}
  {96}},\ \bibinfo {pages} {041103} (\bibinfo {year} {2017})}\BibitemShut
  {NoStop}%
\bibitem [{\citenamefont {Xu}\ \emph {et~al.}(2018)\citenamefont {Xu},
  \citenamefont {Qian}, \citenamefont {Wu}, \citenamefont {Aut\`es},
  \citenamefont {Matt}, \citenamefont {Lv}, \citenamefont {Yao}, \citenamefont
  {Strocov}, \citenamefont {Pomjakushina}, \citenamefont {Conder},
  \citenamefont {Plumb}, \citenamefont {Radovic}, \citenamefont {Yazyev},
  \citenamefont {Qian}, \citenamefont {Ding}, \citenamefont {Mesot},\ and\
  \citenamefont {Shi}}]{PhysRevB.97.161111}%
  \BibitemOpen
  \bibfield  {author} {\bibinfo {author} {\bibfnamefont {N.}~\bibnamefont
  {Xu}}, \bibinfo {author} {\bibfnamefont {Y.~T.}\ \bibnamefont {Qian}},
  \bibinfo {author} {\bibfnamefont {Q.~S.}\ \bibnamefont {Wu}}, \bibinfo
  {author} {\bibfnamefont {G.}~\bibnamefont {Aut\`es}}, \bibinfo {author}
  {\bibfnamefont {C.~E.}\ \bibnamefont {Matt}}, \bibinfo {author}
  {\bibfnamefont {B.~Q.}\ \bibnamefont {Lv}}, \bibinfo {author} {\bibfnamefont
  {M.~Y.}\ \bibnamefont {Yao}}, \bibinfo {author} {\bibfnamefont {V.~N.}\
  \bibnamefont {Strocov}}, \bibinfo {author} {\bibfnamefont {E.}~\bibnamefont
  {Pomjakushina}}, \bibinfo {author} {\bibfnamefont {K.}~\bibnamefont
  {Conder}}, \bibinfo {author} {\bibfnamefont {N.~C.}\ \bibnamefont {Plumb}},
  \bibinfo {author} {\bibfnamefont {M.}~\bibnamefont {Radovic}}, \bibinfo
  {author} {\bibfnamefont {O.~V.}\ \bibnamefont {Yazyev}}, \bibinfo {author}
  {\bibfnamefont {T.}~\bibnamefont {Qian}}, \bibinfo {author} {\bibfnamefont
  {H.}~\bibnamefont {Ding}}, \bibinfo {author} {\bibfnamefont {J.}~\bibnamefont
  {Mesot}},\ and\ \bibinfo {author} {\bibfnamefont {M.}~\bibnamefont {Shi}},\
  }\bibfield  {title} {\bibinfo {title} {Trivial topological phase of
  {C}a{A}g{P} and the topological nodal-line transition in
  {C}a{A}g($\textrm{P}_{1-x}${A}$\textrm{s}_{x}$)},\ }\href
  {https://doi.org/10.1103/PhysRevB.97.161111} {\bibfield  {journal} {\bibinfo
  {journal} {Phys. Rev. B}\ }\textbf {\bibinfo {volume} {97}},\ \bibinfo
  {pages} {161111} (\bibinfo {year} {2018})}\BibitemShut {NoStop}%
\bibitem [{\citenamefont {Wang}\ \emph
  {et~al.}(2020{\natexlab{b}})\citenamefont {Wang}, \citenamefont {Dai},
  \citenamefont {Shao}, \citenamefont {Yang},\ and\ \citenamefont
  {Zhao}}]{PhysRevLett.125.126403}%
  \BibitemOpen
  \bibfield  {author} {\bibinfo {author} {\bibfnamefont {K.}~\bibnamefont
  {Wang}}, \bibinfo {author} {\bibfnamefont {J.-X.}\ \bibnamefont {Dai}},
  \bibinfo {author} {\bibfnamefont {L.~B.}\ \bibnamefont {Shao}}, \bibinfo
  {author} {\bibfnamefont {S.~A.}\ \bibnamefont {Yang}},\ and\ \bibinfo
  {author} {\bibfnamefont {Y.~X.}\ \bibnamefont {Zhao}},\ }\bibfield  {title}
  {\bibinfo {title} {Boundary {C}riticality of $\mathcal{PT}$-{I}nvariant
  {T}opology and {S}econd-{O}rder {N}odal-{L}ine {S}emimetals},\ }\href
  {https://doi.org/10.1103/PhysRevLett.125.126403} {\bibfield  {journal}
  {\bibinfo  {journal} {Phys. Rev. Lett.}\ }\textbf {\bibinfo {volume} {125}},\
  \bibinfo {pages} {126403} (\bibinfo {year} {2020}{\natexlab{b}})}\BibitemShut
  {NoStop}%
\bibitem [{\citenamefont {Qiu}\ \emph {et~al.}(2024)\citenamefont {Qiu},
  \citenamefont {Li}, \citenamefont {Zhang},\ and\ \citenamefont
  {Qiu}}]{PhysRevLett.132.186601}%
  \BibitemOpen
  \bibfield  {author} {\bibinfo {author} {\bibfnamefont {H.}~\bibnamefont
  {Qiu}}, \bibinfo {author} {\bibfnamefont {Y.}~\bibnamefont {Li}}, \bibinfo
  {author} {\bibfnamefont {Q.}~\bibnamefont {Zhang}},\ and\ \bibinfo {author}
  {\bibfnamefont {C.}~\bibnamefont {Qiu}},\ }\bibfield  {title} {\bibinfo
  {title} {Discovery of {H}igher-{O}rder {N}odal {S}urface {S}emimetals},\
  }\href {https://doi.org/10.1103/PhysRevLett.132.186601} {\bibfield  {journal}
  {\bibinfo  {journal} {Phys. Rev. Lett.}\ }\textbf {\bibinfo {volume} {132}},\
  \bibinfo {pages} {186601} (\bibinfo {year} {2024})}\BibitemShut {NoStop}%
\bibitem [{\citenamefont {Li}\ \emph {et~al.}(2021)\citenamefont {Li},
  \citenamefont {Wang},\ and\ \citenamefont {Pan}}]{PhysRevB.104.235136}%
  \BibitemOpen
  \bibfield  {author} {\bibinfo {author} {\bibfnamefont {J.}~\bibnamefont
  {Li}}, \bibinfo {author} {\bibfnamefont {H.}~\bibnamefont {Wang}},\ and\
  \bibinfo {author} {\bibfnamefont {H.}~\bibnamefont {Pan}},\ }\bibfield
  {title} {\bibinfo {title} {Tunable topological phase transition from
  nodal-line semimetal to {W}eyl semimetal by breaking symmetry},\ }\href
  {https://doi.org/10.1103/PhysRevB.104.235136} {\bibfield  {journal} {\bibinfo
   {journal} {Phys. Rev. B}\ }\textbf {\bibinfo {volume} {104}},\ \bibinfo
  {pages} {235136} (\bibinfo {year} {2021})}\BibitemShut {NoStop}%
\bibitem [{\citenamefont {Yan}\ and\ \citenamefont
  {Wang}(2016)}]{PhysRevLett.117.087402}%
  \BibitemOpen
  \bibfield  {author} {\bibinfo {author} {\bibfnamefont {Z.}~\bibnamefont
  {Yan}}\ and\ \bibinfo {author} {\bibfnamefont {Z.}~\bibnamefont {Wang}},\
  }\bibfield  {title} {\bibinfo {title} {Tunable {W}eyl {P}oints in
  {P}eriodically {D}riven {N}odal {L}ine {S}emimetals},\ }\href
  {https://doi.org/10.1103/PhysRevLett.117.087402} {\bibfield  {journal}
  {\bibinfo  {journal} {Phys. Rev. Lett.}\ }\textbf {\bibinfo {volume} {117}},\
  \bibinfo {pages} {087402} (\bibinfo {year} {2016})}\BibitemShut {NoStop}%
\bibitem [{\citenamefont {Yan}\ and\ \citenamefont
  {Wang}(2017)}]{PhysRevB.96.041206}%
  \BibitemOpen
  \bibfield  {author} {\bibinfo {author} {\bibfnamefont {Z.}~\bibnamefont
  {Yan}}\ and\ \bibinfo {author} {\bibfnamefont {Z.}~\bibnamefont {Wang}},\
  }\bibfield  {title} {\bibinfo {title} {Floquet multi-{W}eyl points in
  crossing-nodal-line semimetals},\ }\href
  {https://doi.org/10.1103/PhysRevB.96.041206} {\bibfield  {journal} {\bibinfo
  {journal} {Phys. Rev. B}\ }\textbf {\bibinfo {volume} {96}},\ \bibinfo
  {pages} {041206} (\bibinfo {year} {2017})}\BibitemShut {NoStop}%
\bibitem [{\citenamefont {Wang}\ \emph {et~al.}(2023)\citenamefont {Wang},
  \citenamefont {Wang}, \citenamefont {Sun}, \citenamefont {Chen},\ and\
  \citenamefont {Xu}}]{PhysRevB.107.L121407}%
  \BibitemOpen
  \bibfield  {author} {\bibinfo {author} {\bibfnamefont {Z.-M.}\ \bibnamefont
  {Wang}}, \bibinfo {author} {\bibfnamefont {R.}~\bibnamefont {Wang}}, \bibinfo
  {author} {\bibfnamefont {J.-H.}\ \bibnamefont {Sun}}, \bibinfo {author}
  {\bibfnamefont {T.-Y.}\ \bibnamefont {Chen}},\ and\ \bibinfo {author}
  {\bibfnamefont {D.-H.}\ \bibnamefont {Xu}},\ }\bibfield  {title} {\bibinfo
  {title} {Floquet weyl semimetal phases in light-irradiated higher-order
  topological dirac semimetals},\ }\href
  {https://doi.org/10.1103/PhysRevB.107.L121407} {\bibfield  {journal}
  {\bibinfo  {journal} {Phys. Rev. B}\ }\textbf {\bibinfo {volume} {107}},\
  \bibinfo {pages} {L121407} (\bibinfo {year} {2023})}\BibitemShut {NoStop}%
\bibitem [{\citenamefont {Fang}\ \emph {et~al.}(2016)\citenamefont {Fang},
  \citenamefont {Weng}, \citenamefont {Dai},\ and\ \citenamefont
  {Fang}}]{Fang_2016}%
  \BibitemOpen
  \bibfield  {author} {\bibinfo {author} {\bibfnamefont {C.}~\bibnamefont
  {Fang}}, \bibinfo {author} {\bibfnamefont {H.}~\bibnamefont {Weng}}, \bibinfo
  {author} {\bibfnamefont {X.}~\bibnamefont {Dai}},\ and\ \bibinfo {author}
  {\bibfnamefont {Z.}~\bibnamefont {Fang}},\ }\bibfield  {title} {\bibinfo
  {title} {Topological nodal line semimetals},\ }\href
  {https://doi.org/10.1088/1674-1056/25/11/117106} {\bibfield  {journal}
  {\bibinfo  {journal} {Chinese Physics B}\ }\textbf {\bibinfo {volume} {25}},\
  \bibinfo {pages} {117106} (\bibinfo {year} {2016})}\BibitemShut {NoStop}%
\bibitem [{\citenamefont {Ezawa}(2016)}]{PhysRevLett.116.127202}%
  \BibitemOpen
  \bibfield  {author} {\bibinfo {author} {\bibfnamefont {M.}~\bibnamefont
  {Ezawa}},\ }\bibfield  {title} {\bibinfo {title} {Loop-{N}odal and
  {P}oint-{N}odal {S}emimetals in {T}hree-{D}imensional {H}oneycomb
  {L}attices},\ }\href {https://doi.org/10.1103/PhysRevLett.116.127202}
  {\bibfield  {journal} {\bibinfo  {journal} {Phys. Rev. Lett.}\ }\textbf
  {\bibinfo {volume} {116}},\ \bibinfo {pages} {127202} (\bibinfo {year}
  {2016})}\BibitemShut {NoStop}%
\bibitem [{\citenamefont {Chen}\ \emph {et~al.}(2019)\citenamefont {Chen},
  \citenamefont {Lu},\ and\ \citenamefont
  {Zilberberg}}]{PhysRevLett.122.196603}%
  \BibitemOpen
  \bibfield  {author} {\bibinfo {author} {\bibfnamefont {W.}~\bibnamefont
  {Chen}}, \bibinfo {author} {\bibfnamefont {H.-Z.}\ \bibnamefont {Lu}},\ and\
  \bibinfo {author} {\bibfnamefont {O.}~\bibnamefont {Zilberberg}},\ }\bibfield
   {title} {\bibinfo {title} {Weak {L}ocalization and {A}ntilocalization in
  {N}odal-{L}ine {S}emimetals: {D}imensionality and {T}opological {E}ffects},\
  }\href {https://doi.org/10.1103/PhysRevLett.122.196603} {\bibfield  {journal}
  {\bibinfo  {journal} {Phys. Rev. Lett.}\ }\textbf {\bibinfo {volume} {122}},\
  \bibinfo {pages} {196603} (\bibinfo {year} {2019})}\BibitemShut {NoStop}%
\bibitem [{\citenamefont {Song}\ \emph {et~al.}(2020)\citenamefont {Song},
  \citenamefont {Wang}, \citenamefont {Li}, \citenamefont {Liu}, \citenamefont
  {Lu}, \citenamefont {Liu}, \citenamefont {Li}, \citenamefont {Wen},
  \citenamefont {Yin}, \citenamefont {Liu},\ and\ \citenamefont
  {Shen}}]{PhysRevLett.124.056402}%
  \BibitemOpen
  \bibfield  {author} {\bibinfo {author} {\bibfnamefont {Y.~K.}\ \bibnamefont
  {Song}}, \bibinfo {author} {\bibfnamefont {G.~W.}\ \bibnamefont {Wang}},
  \bibinfo {author} {\bibfnamefont {S.~C.}\ \bibnamefont {Li}}, \bibinfo
  {author} {\bibfnamefont {W.~L.}\ \bibnamefont {Liu}}, \bibinfo {author}
  {\bibfnamefont {X.~L.}\ \bibnamefont {Lu}}, \bibinfo {author} {\bibfnamefont
  {Z.~T.}\ \bibnamefont {Liu}}, \bibinfo {author} {\bibfnamefont {Z.~J.}\
  \bibnamefont {Li}}, \bibinfo {author} {\bibfnamefont {J.~S.}\ \bibnamefont
  {Wen}}, \bibinfo {author} {\bibfnamefont {Z.~P.}\ \bibnamefont {Yin}},
  \bibinfo {author} {\bibfnamefont {Z.~H.}\ \bibnamefont {Liu}},\ and\ \bibinfo
  {author} {\bibfnamefont {D.~W.}\ \bibnamefont {Shen}},\ }\bibfield  {title}
  {\bibinfo {title} {{P}hotoemission {S}pectroscopic {E}vidence for the {D}irac
  {N}odal {L}ine in the {M}onoclinic {S}emimetal ${\mathrm{sras}}_{3}$},\
  }\href {https://doi.org/10.1103/PhysRevLett.124.056402} {\bibfield  {journal}
  {\bibinfo  {journal} {Phys. Rev. Lett.}\ }\textbf {\bibinfo {volume} {124}},\
  \bibinfo {pages} {056402} (\bibinfo {year} {2020})}\BibitemShut {NoStop}%
\bibitem [{\citenamefont {Son}\ and\ \citenamefont
  {Spivak}(2013)}]{PhysRevB.88.104412}%
  \BibitemOpen
  \bibfield  {author} {\bibinfo {author} {\bibfnamefont {D.~T.}\ \bibnamefont
  {Son}}\ and\ \bibinfo {author} {\bibfnamefont {B.~Z.}\ \bibnamefont
  {Spivak}},\ }\bibfield  {title} {\bibinfo {title} {Chiral anomaly and
  classical negative magnetoresistance of {W}eyl metals},\ }\href
  {https://doi.org/10.1103/PhysRevB.88.104412} {\bibfield  {journal} {\bibinfo
  {journal} {Phys. Rev. B}\ }\textbf {\bibinfo {volume} {88}},\ \bibinfo
  {pages} {104412} (\bibinfo {year} {2013})}\BibitemShut {NoStop}%
\bibitem [{\citenamefont {Cano}\ \emph {et~al.}(2017)\citenamefont {Cano},
  \citenamefont {Bradlyn}, \citenamefont {Wang}, \citenamefont {Hirschberger},
  \citenamefont {Ong},\ and\ \citenamefont {Bernevig}}]{PhysRevB.95.161306}%
  \BibitemOpen
  \bibfield  {author} {\bibinfo {author} {\bibfnamefont {J.}~\bibnamefont
  {Cano}}, \bibinfo {author} {\bibfnamefont {B.}~\bibnamefont {Bradlyn}},
  \bibinfo {author} {\bibfnamefont {Z.}~\bibnamefont {Wang}}, \bibinfo {author}
  {\bibfnamefont {M.}~\bibnamefont {Hirschberger}}, \bibinfo {author}
  {\bibfnamefont {N.~P.}\ \bibnamefont {Ong}},\ and\ \bibinfo {author}
  {\bibfnamefont {B.~A.}\ \bibnamefont {Bernevig}},\ }\bibfield  {title}
  {\bibinfo {title} {Chiral anomaly factory: {C}reating {W}eyl fermions with a
  magnetic field},\ }\href {https://doi.org/10.1103/PhysRevB.95.161306}
  {\bibfield  {journal} {\bibinfo  {journal} {Phys. Rev. B}\ }\textbf {\bibinfo
  {volume} {95}},\ \bibinfo {pages} {161306} (\bibinfo {year}
  {2017})}\BibitemShut {NoStop}%
\bibitem [{\citenamefont {Zyuzin}\ and\ \citenamefont
  {Burkov}(2012)}]{PhysRevB.86.115133}%
  \BibitemOpen
  \bibfield  {author} {\bibinfo {author} {\bibfnamefont {A.~A.}\ \bibnamefont
  {Zyuzin}}\ and\ \bibinfo {author} {\bibfnamefont {A.~A.}\ \bibnamefont
  {Burkov}},\ }\bibfield  {title} {\bibinfo {title} {Topological response in
  {W}eyl semimetals and the chiral anomaly},\ }\href
  {https://doi.org/10.1103/PhysRevB.86.115133} {\bibfield  {journal} {\bibinfo
  {journal} {Phys. Rev. B}\ }\textbf {\bibinfo {volume} {86}},\ \bibinfo
  {pages} {115133} (\bibinfo {year} {2012})}\BibitemShut {NoStop}%
\bibitem [{\citenamefont {Vazifeh}\ and\ \citenamefont
  {Franz}(2013)}]{PhysRevLett.111.027201}%
  \BibitemOpen
  \bibfield  {author} {\bibinfo {author} {\bibfnamefont {M.~M.}\ \bibnamefont
  {Vazifeh}}\ and\ \bibinfo {author} {\bibfnamefont {M.}~\bibnamefont
  {Franz}},\ }\bibfield  {title} {\bibinfo {title} {Electromagnetic response of
  {W}eyl semimetals},\ }\href {https://doi.org/10.1103/PhysRevLett.111.027201}
  {\bibfield  {journal} {\bibinfo  {journal} {Phys. Rev. Lett.}\ }\textbf
  {\bibinfo {volume} {111}},\ \bibinfo {pages} {027201} (\bibinfo {year}
  {2013})}\BibitemShut {NoStop}%
\bibitem [{\citenamefont {Huang}\ \emph {et~al.}(2015)\citenamefont {Huang},
  \citenamefont {Zhao}, \citenamefont {Long}, \citenamefont {Wang},
  \citenamefont {Chen}, \citenamefont {Yang}, \citenamefont {Liang},
  \citenamefont {Xue}, \citenamefont {Weng}, \citenamefont {Fang},
  \citenamefont {Dai},\ and\ \citenamefont {Chen}}]{PhysRevX.5.031023}%
  \BibitemOpen
  \bibfield  {author} {\bibinfo {author} {\bibfnamefont {X.}~\bibnamefont
  {Huang}}, \bibinfo {author} {\bibfnamefont {L.}~\bibnamefont {Zhao}},
  \bibinfo {author} {\bibfnamefont {Y.}~\bibnamefont {Long}}, \bibinfo {author}
  {\bibfnamefont {P.}~\bibnamefont {Wang}}, \bibinfo {author} {\bibfnamefont
  {D.}~\bibnamefont {Chen}}, \bibinfo {author} {\bibfnamefont {Z.}~\bibnamefont
  {Yang}}, \bibinfo {author} {\bibfnamefont {H.}~\bibnamefont {Liang}},
  \bibinfo {author} {\bibfnamefont {M.}~\bibnamefont {Xue}}, \bibinfo {author}
  {\bibfnamefont {H.}~\bibnamefont {Weng}}, \bibinfo {author} {\bibfnamefont
  {Z.}~\bibnamefont {Fang}}, \bibinfo {author} {\bibfnamefont {X.}~\bibnamefont
  {Dai}},\ and\ \bibinfo {author} {\bibfnamefont {G.}~\bibnamefont {Chen}},\
  }\bibfield  {title} {\bibinfo {title} {Observation of the
  {C}hiral-{A}nomaly-{I}nduced {N}egative {M}agnetoresistance in 3{D} {W}eyl
  {S}emimetal {T}a{A}s},\ }\href {https://doi.org/10.1103/PhysRevX.5.031023}
  {\bibfield  {journal} {\bibinfo  {journal} {Phys. Rev. X}\ }\textbf {\bibinfo
  {volume} {5}},\ \bibinfo {pages} {031023} (\bibinfo {year}
  {2015})}\BibitemShut {NoStop}%
\bibitem [{\citenamefont {Gorbar}\ \emph {et~al.}(2014)\citenamefont {Gorbar},
  \citenamefont {Miransky},\ and\ \citenamefont
  {Shovkovy}}]{PhysRevB.89.085126}%
  \BibitemOpen
  \bibfield  {author} {\bibinfo {author} {\bibfnamefont {E.~V.}\ \bibnamefont
  {Gorbar}}, \bibinfo {author} {\bibfnamefont {V.~A.}\ \bibnamefont
  {Miransky}},\ and\ \bibinfo {author} {\bibfnamefont {I.~A.}\ \bibnamefont
  {Shovkovy}},\ }\bibfield  {title} {\bibinfo {title} {Chiral anomaly,
  dimensional reduction, and magnetoresistivity of {W}eyl and {D}irac
  semimetals},\ }\href {https://doi.org/10.1103/PhysRevB.89.085126} {\bibfield
  {journal} {\bibinfo  {journal} {Phys. Rev. B}\ }\textbf {\bibinfo {volume}
  {89}},\ \bibinfo {pages} {085126} (\bibinfo {year} {2014})}\BibitemShut
  {NoStop}%
\bibitem [{\citenamefont {Liang}\ \emph {et~al.}(2015)\citenamefont {Liang},
  \citenamefont {Gibson}, \citenamefont {Ali}, \citenamefont {Liu},
  \citenamefont {Cava},\ and\ \citenamefont {Ong}}]{LiangTian2015}%
  \BibitemOpen
  \bibfield  {author} {\bibinfo {author} {\bibfnamefont {T.}~\bibnamefont
  {Liang}}, \bibinfo {author} {\bibfnamefont {Q.}~\bibnamefont {Gibson}},
  \bibinfo {author} {\bibfnamefont {M.~N.}\ \bibnamefont {Ali}}, \bibinfo
  {author} {\bibfnamefont {M.}~\bibnamefont {Liu}}, \bibinfo {author}
  {\bibfnamefont {R.~J.}\ \bibnamefont {Cava}},\ and\ \bibinfo {author}
  {\bibfnamefont {N.~P.}\ \bibnamefont {Ong}},\ }\bibfield  {title} {\bibinfo
  {title} {Ultrahigh mobility and giant magnetoresistance in the {D}irac
  semimetal {C}d3{A}s2},\ }\href {https://doi.org/10.1038/nmat4143} {\bibfield
   {journal} {\bibinfo  {journal} {Nature Materials}\ }\textbf {\bibinfo
  {volume} {14}},\ \bibinfo {pages} {280} (\bibinfo {year} {2015})}\BibitemShut
  {NoStop}%
\bibitem [{\citenamefont {Chen}\ \emph {et~al.}(2018)\citenamefont {Chen},
  \citenamefont {Luo}, \citenamefont {Li},\ and\ \citenamefont
  {Zilberberg}}]{PhysRevLett.121.166802}%
  \BibitemOpen
  \bibfield  {author} {\bibinfo {author} {\bibfnamefont {W.}~\bibnamefont
  {Chen}}, \bibinfo {author} {\bibfnamefont {K.}~\bibnamefont {Luo}}, \bibinfo
  {author} {\bibfnamefont {L.}~\bibnamefont {Li}},\ and\ \bibinfo {author}
  {\bibfnamefont {O.}~\bibnamefont {Zilberberg}},\ }\bibfield  {title}
  {\bibinfo {title} {Proposal for {D}etecting {N}odal-{L}ine {S}emimetal
  {S}urface {S}tates with {R}esonant {S}pin-{F}lipped {R}eflection},\ }\href
  {https://doi.org/10.1103/PhysRevLett.121.166802} {\bibfield  {journal}
  {\bibinfo  {journal} {Phys. Rev. Lett.}\ }\textbf {\bibinfo {volume} {121}},\
  \bibinfo {pages} {166802} (\bibinfo {year} {2018})}\BibitemShut {NoStop}%
\bibitem [{\citenamefont {Li}\ \emph {et~al.}(2018{\natexlab{a}})\citenamefont
  {Li}, \citenamefont {Wang}, \citenamefont {Wan}, \citenamefont {Wan},
  \citenamefont {Lu},\ and\ \citenamefont {Xie}}]{PhysRevLett.120.146602}%
  \BibitemOpen
  \bibfield  {author} {\bibinfo {author} {\bibfnamefont {C.}~\bibnamefont
  {Li}}, \bibinfo {author} {\bibfnamefont {C.~M.}\ \bibnamefont {Wang}},
  \bibinfo {author} {\bibfnamefont {B.}~\bibnamefont {Wan}}, \bibinfo {author}
  {\bibfnamefont {X.}~\bibnamefont {Wan}}, \bibinfo {author} {\bibfnamefont
  {H.-Z.}\ \bibnamefont {Lu}},\ and\ \bibinfo {author} {\bibfnamefont {X.~C.}\
  \bibnamefont {Xie}},\ }\bibfield  {title} {\bibinfo {title} {Rules for
  {P}hase {S}hifts of {Q}uantum {O}scillations in {T}opological {N}odal-{L}ine
  {S}emimetals},\ }\href {https://doi.org/10.1103/PhysRevLett.120.146602}
  {\bibfield  {journal} {\bibinfo  {journal} {Phys. Rev. Lett.}\ }\textbf
  {\bibinfo {volume} {120}},\ \bibinfo {pages} {146602} (\bibinfo {year}
  {2018}{\natexlab{a}})}\BibitemShut {NoStop}%
\bibitem [{\citenamefont {Hu}\ \emph {et~al.}(2024)\citenamefont {Hu},
  \citenamefont {Guo}, \citenamefont {Liu}, \citenamefont {Chen},\ and\
  \citenamefont {Chen}}]{Hu2024}%
  \BibitemOpen
  \bibfield  {author} {\bibinfo {author} {\bibfnamefont {S.}~\bibnamefont
  {Hu}}, \bibinfo {author} {\bibfnamefont {Z.}~\bibnamefont {Guo}}, \bibinfo
  {author} {\bibfnamefont {W.}~\bibnamefont {Liu}}, \bibinfo {author}
  {\bibfnamefont {S.}~\bibnamefont {Chen}},\ and\ \bibinfo {author}
  {\bibfnamefont {H.}~\bibnamefont {Chen}},\ }\bibfield  {title} {\bibinfo
  {title} {Hyperbolic metamaterial empowered controllable photonic weyl nodal
  line semimetals},\ }\href {https://doi.org/10.1038/s41467-024-47125-7}
  {\bibfield  {journal} {\bibinfo  {journal} {Nature Communications}\ }\textbf
  {\bibinfo {volume} {15}},\ \bibinfo {pages} {2773} (\bibinfo {year}
  {2024})}\BibitemShut {NoStop}%
\bibitem [{\citenamefont {Cheng}\ \emph {et~al.}(2020)\citenamefont {Cheng},
  \citenamefont {Gao}, \citenamefont {Bi}, \citenamefont {Liu}, \citenamefont
  {Li}, \citenamefont {Guo}, \citenamefont {Yang}, \citenamefont {You},
  \citenamefont {Feng}, \citenamefont {Sun}, \citenamefont {Tian},
  \citenamefont {Chen},\ and\ \citenamefont {Zhang}}]{PhysRevLett.125.093904}%
  \BibitemOpen
  \bibfield  {author} {\bibinfo {author} {\bibfnamefont {H.}~\bibnamefont
  {Cheng}}, \bibinfo {author} {\bibfnamefont {W.}~\bibnamefont {Gao}}, \bibinfo
  {author} {\bibfnamefont {Y.}~\bibnamefont {Bi}}, \bibinfo {author}
  {\bibfnamefont {W.}~\bibnamefont {Liu}}, \bibinfo {author} {\bibfnamefont
  {Z.}~\bibnamefont {Li}}, \bibinfo {author} {\bibfnamefont {Q.}~\bibnamefont
  {Guo}}, \bibinfo {author} {\bibfnamefont {Y.}~\bibnamefont {Yang}}, \bibinfo
  {author} {\bibfnamefont {O.}~\bibnamefont {You}}, \bibinfo {author}
  {\bibfnamefont {J.}~\bibnamefont {Feng}}, \bibinfo {author} {\bibfnamefont
  {H.}~\bibnamefont {Sun}}, \bibinfo {author} {\bibfnamefont {J.}~\bibnamefont
  {Tian}}, \bibinfo {author} {\bibfnamefont {S.}~\bibnamefont {Chen}},\ and\
  \bibinfo {author} {\bibfnamefont {S.}~\bibnamefont {Zhang}},\ }\bibfield
  {title} {\bibinfo {title} {Vortical reflection and spiraling fermi arcs with
  weyl metamaterials},\ }\href {https://doi.org/10.1103/PhysRevLett.125.093904}
  {\bibfield  {journal} {\bibinfo  {journal} {Phys. Rev. Lett.}\ }\textbf
  {\bibinfo {volume} {125}},\ \bibinfo {pages} {093904} (\bibinfo {year}
  {2020})}\BibitemShut {NoStop}%
\bibitem [{\citenamefont {Yang}\ \emph
  {et~al.}(2024{\natexlab{a}})\citenamefont {Yang}, \citenamefont {Song},
  \citenamefont {Cao},\ and\ \citenamefont {Yan}}]{YANG20241}%
  \BibitemOpen
  \bibfield  {author} {\bibinfo {author} {\bibfnamefont {H.}~\bibnamefont
  {Yang}}, \bibinfo {author} {\bibfnamefont {L.}~\bibnamefont {Song}}, \bibinfo
  {author} {\bibfnamefont {Y.}~\bibnamefont {Cao}},\ and\ \bibinfo {author}
  {\bibfnamefont {P.}~\bibnamefont {Yan}},\ }\bibfield  {title} {\bibinfo
  {title} {Circuit realization of topological physics},\ }\href
  {https://doi.org/https://doi.org/10.1016/j.physrep.2024.09.007} {\bibfield
  {journal} {\bibinfo  {journal} {Physics Reports}\ }\textbf {\bibinfo {volume}
  {1093}},\ \bibinfo {pages} {1} (\bibinfo {year}
  {2024}{\natexlab{a}})}\BibitemShut {NoStop}%
\bibitem [{\citenamefont {Dong}\ \emph {et~al.}(2021)\citenamefont {Dong},
  \citenamefont {Juri\ifmmode \check{c}\else \v{c}\fi{}i\ifmmode~\acute{c}\else
  \'{c}\fi{}},\ and\ \citenamefont {Roy}}]{PhysRevResearch.3.023056}%
  \BibitemOpen
  \bibfield  {author} {\bibinfo {author} {\bibfnamefont {J.}~\bibnamefont
  {Dong}}, \bibinfo {author} {\bibfnamefont {V.}~\bibnamefont {Juri\ifmmode
  \check{c}\else \v{c}\fi{}i\ifmmode~\acute{c}\else \'{c}\fi{}}},\ and\
  \bibinfo {author} {\bibfnamefont {B.}~\bibnamefont {Roy}},\ }\bibfield
  {title} {\bibinfo {title} {Topolectric circuits: Theory and construction},\
  }\href {https://doi.org/10.1103/PhysRevResearch.3.023056} {\bibfield
  {journal} {\bibinfo  {journal} {Phys. Rev. Res.}\ }\textbf {\bibinfo {volume}
  {3}},\ \bibinfo {pages} {023056} (\bibinfo {year} {2021})}\BibitemShut
  {NoStop}%
\bibitem [{\citenamefont {Rafi-Ul-Islam}\ \emph {et~al.}(2020)\citenamefont
  {Rafi-Ul-Islam}, \citenamefont {Bin~Siu},\ and\ \citenamefont
  {Jalil}}]{Rafi-Ul-Islam2020}%
  \BibitemOpen
  \bibfield  {author} {\bibinfo {author} {\bibfnamefont {S.~M.}\ \bibnamefont
  {Rafi-Ul-Islam}}, \bibinfo {author} {\bibfnamefont {Z.}~\bibnamefont
  {Bin~Siu}},\ and\ \bibinfo {author} {\bibfnamefont {M.~B.~A.}\ \bibnamefont
  {Jalil}},\ }\bibfield  {title} {\bibinfo {title} {Topoelectrical circuit
  realization of a weyl semimetal heterojunction},\ }\href
  {https://doi.org/10.1038/s42005-020-0336-0} {\bibfield  {journal} {\bibinfo
  {journal} {Communications Physics}\ }\textbf {\bibinfo {volume} {3}},\
  \bibinfo {pages} {72} (\bibinfo {year} {2020})}\BibitemShut {NoStop}%
\bibitem [{\citenamefont {Yang}\ \emph {et~al.}(2021)\citenamefont {Yang},
  \citenamefont {Lu}, \citenamefont {Yan}, \citenamefont {Huang}, \citenamefont
  {Deng},\ and\ \citenamefont {Liu}}]{PhysRevLett.126.156801}%
  \BibitemOpen
  \bibfield  {author} {\bibinfo {author} {\bibfnamefont {Y.}~\bibnamefont
  {Yang}}, \bibinfo {author} {\bibfnamefont {J.}~\bibnamefont {Lu}}, \bibinfo
  {author} {\bibfnamefont {M.}~\bibnamefont {Yan}}, \bibinfo {author}
  {\bibfnamefont {X.}~\bibnamefont {Huang}}, \bibinfo {author} {\bibfnamefont
  {W.}~\bibnamefont {Deng}},\ and\ \bibinfo {author} {\bibfnamefont
  {Z.}~\bibnamefont {Liu}},\ }\bibfield  {title} {\bibinfo {title}
  {Hybrid-order topological insulators in a phononic crystal},\ }\href
  {https://doi.org/10.1103/PhysRevLett.126.156801} {\bibfield  {journal}
  {\bibinfo  {journal} {Phys. Rev. Lett.}\ }\textbf {\bibinfo {volume} {126}},\
  \bibinfo {pages} {156801} (\bibinfo {year} {2021})}\BibitemShut {NoStop}%
\bibitem [{\citenamefont {Liu}\ \emph {et~al.}(2024)\citenamefont {Liu},
  \citenamefont {Ding}, \citenamefont {Luo}, \citenamefont {Zeng},
  \citenamefont {Tang},\ and\ \citenamefont {Chen}}]{PhysRevB.109.115422}%
  \BibitemOpen
  \bibfield  {author} {\bibinfo {author} {\bibfnamefont {W.}~\bibnamefont
  {Liu}}, \bibinfo {author} {\bibfnamefont {Z.-K.}\ \bibnamefont {Ding}},
  \bibinfo {author} {\bibfnamefont {N.}~\bibnamefont {Luo}}, \bibinfo {author}
  {\bibfnamefont {J.}~\bibnamefont {Zeng}}, \bibinfo {author} {\bibfnamefont
  {L.-M.}\ \bibnamefont {Tang}},\ and\ \bibinfo {author} {\bibfnamefont
  {K.-Q.}\ \bibnamefont {Chen}},\ }\bibfield  {title} {\bibinfo {title}
  {Phononic hybrid-order topology in semihydrogenated graphene},\ }\href
  {https://doi.org/10.1103/PhysRevB.109.115422} {\bibfield  {journal} {\bibinfo
   {journal} {Phys. Rev. B}\ }\textbf {\bibinfo {volume} {109}},\ \bibinfo
  {pages} {115422} (\bibinfo {year} {2024})}\BibitemShut {NoStop}%
\bibitem [{\citenamefont {Wu}\ and\ \citenamefont
  {An}(2023)}]{PhysRevB.107.235132}%
  \BibitemOpen
  \bibfield  {author} {\bibinfo {author} {\bibfnamefont {H.}~\bibnamefont
  {Wu}}\ and\ \bibinfo {author} {\bibfnamefont {J.-H.}\ \bibnamefont {An}},\
  }\bibfield  {title} {\bibinfo {title} {Hybrid-order topological odd-parity
  superconductors via floquet engineering},\ }\href
  {https://doi.org/10.1103/PhysRevB.107.235132} {\bibfield  {journal} {\bibinfo
   {journal} {Phys. Rev. B}\ }\textbf {\bibinfo {volume} {107}},\ \bibinfo
  {pages} {235132} (\bibinfo {year} {2023})}\BibitemShut {NoStop}%
\bibitem [{\citenamefont {Lai}\ \emph {et~al.}(2024)\citenamefont {Lai},
  \citenamefont {Wu}, \citenamefont {Pu}, \citenamefont {Zhou}, \citenamefont
  {Lu}, \citenamefont {Liu}, \citenamefont {Deng}, \citenamefont {Cheng},
  \citenamefont {Chen},\ and\ \citenamefont {Liu}}]{PhysRevApplied.21.044002}%
  \BibitemOpen
  \bibfield  {author} {\bibinfo {author} {\bibfnamefont {P.}~\bibnamefont
  {Lai}}, \bibinfo {author} {\bibfnamefont {J.}~\bibnamefont {Wu}}, \bibinfo
  {author} {\bibfnamefont {Z.}~\bibnamefont {Pu}}, \bibinfo {author}
  {\bibfnamefont {Q.}~\bibnamefont {Zhou}}, \bibinfo {author} {\bibfnamefont
  {J.}~\bibnamefont {Lu}}, \bibinfo {author} {\bibfnamefont {H.}~\bibnamefont
  {Liu}}, \bibinfo {author} {\bibfnamefont {W.}~\bibnamefont {Deng}}, \bibinfo
  {author} {\bibfnamefont {H.}~\bibnamefont {Cheng}}, \bibinfo {author}
  {\bibfnamefont {S.}~\bibnamefont {Chen}},\ and\ \bibinfo {author}
  {\bibfnamefont {Z.}~\bibnamefont {Liu}},\ }\bibfield  {title} {\bibinfo
  {title} {Real-projective-plane hybrid-order topological insulator realized in
  phononic crystals},\ }\href
  {https://doi.org/10.1103/PhysRevApplied.21.044002} {\bibfield  {journal}
  {\bibinfo  {journal} {Phys. Rev. Appl.}\ }\textbf {\bibinfo {volume} {21}},\
  \bibinfo {pages} {044002} (\bibinfo {year} {2024})}\BibitemShut {NoStop}%
\bibitem [{\citenamefont {Huang}\ \emph {et~al.}(2024)\citenamefont {Huang},
  \citenamefont {Park},\ and\ \citenamefont {Hsu}}]{HuangShengJie2024}%
  \BibitemOpen
  \bibfield  {author} {\bibinfo {author} {\bibfnamefont {S.-J.}\ \bibnamefont
  {Huang}}, \bibinfo {author} {\bibfnamefont {K.}~\bibnamefont {Park}},\ and\
  \bibinfo {author} {\bibfnamefont {Y.-T.}\ \bibnamefont {Hsu}},\ }\bibfield
  {title} {\bibinfo {title} {Hybrid-order topological superconductivity in a
  topological metal 1t’-mote2},\ }\href
  {https://doi.org/10.1038/s41535-024-00633-7} {\bibfield  {journal} {\bibinfo
  {journal} {npj Quantum Materials}\ }\textbf {\bibinfo {volume} {9}},\
  \bibinfo {pages} {21} (\bibinfo {year} {2024})}\BibitemShut {NoStop}%
\bibitem [{\citenamefont {Xia}\ \emph {et~al.}(2023)\citenamefont {Xia},
  \citenamefont {Li}, \citenamefont {Zhang}, \citenamefont {Fan}, \citenamefont
  {Xiao},\ and\ \citenamefont {Qiu}}]{PhysRevB.108.125125}%
  \BibitemOpen
  \bibfield  {author} {\bibinfo {author} {\bibfnamefont {T.}~\bibnamefont
  {Xia}}, \bibinfo {author} {\bibfnamefont {Y.}~\bibnamefont {Li}}, \bibinfo
  {author} {\bibfnamefont {Q.}~\bibnamefont {Zhang}}, \bibinfo {author}
  {\bibfnamefont {X.}~\bibnamefont {Fan}}, \bibinfo {author} {\bibfnamefont
  {M.}~\bibnamefont {Xiao}},\ and\ \bibinfo {author} {\bibfnamefont
  {C.}~\bibnamefont {Qiu}},\ }\bibfield  {title} {\bibinfo {title} {Observation
  of hybrid-order topological pump in a kekul\'e-textured graphene lattice},\
  }\href {https://doi.org/10.1103/PhysRevB.108.125125} {\bibfield  {journal}
  {\bibinfo  {journal} {Phys. Rev. B}\ }\textbf {\bibinfo {volume} {108}},\
  \bibinfo {pages} {125125} (\bibinfo {year} {2023})}\BibitemShut {NoStop}%
\bibitem [{\citenamefont {Yang}\ \emph
  {et~al.}(2024{\natexlab{b}})\citenamefont {Yang}, \citenamefont {Huang},\
  and\ \citenamefont {Zhang}}]{10.1063/5.0238775}%
  \BibitemOpen
  \bibfield  {author} {\bibinfo {author} {\bibfnamefont {N.-J.}\ \bibnamefont
  {Yang}}, \bibinfo {author} {\bibfnamefont {Z.}~\bibnamefont {Huang}},\ and\
  \bibinfo {author} {\bibfnamefont {J.-M.}\ \bibnamefont {Zhang}},\ }\bibfield
  {title} {\bibinfo {title} {Hybrid-order topological phase and transition in
  1h transition metal compounds},\ }\href {https://doi.org/10.1063/5.0238775}
  {\bibfield  {journal} {\bibinfo  {journal} {Applied Physics Letters}\
  }\textbf {\bibinfo {volume} {125}},\ \bibinfo {pages} {263102} (\bibinfo
  {year} {2024}{\natexlab{b}})}\BibitemShut {NoStop}%
\bibitem [{\citenamefont {Sun}\ \emph {et~al.}(2024)\citenamefont {Sun},
  \citenamefont {Luo}, \citenamefont {Huang}, \citenamefont {Peng},
  \citenamefont {Zhao}, \citenamefont {Yao}, \citenamefont {Wu},\ and\
  \citenamefont {Zhang}}]{PhysRevB.109.134302}%
  \BibitemOpen
  \bibfield  {author} {\bibinfo {author} {\bibfnamefont {W.}~\bibnamefont
  {Sun}}, \bibinfo {author} {\bibfnamefont {L.}~\bibnamefont {Luo}}, \bibinfo
  {author} {\bibfnamefont {Y.}~\bibnamefont {Huang}}, \bibinfo {author}
  {\bibfnamefont {J.}~\bibnamefont {Peng}}, \bibinfo {author} {\bibfnamefont
  {D.}~\bibnamefont {Zhao}}, \bibinfo {author} {\bibfnamefont {Y.}~\bibnamefont
  {Yao}}, \bibinfo {author} {\bibfnamefont {F.}~\bibnamefont {Wu}},\ and\
  \bibinfo {author} {\bibfnamefont {X.}~\bibnamefont {Zhang}},\ }\bibfield
  {title} {\bibinfo {title} {Observation of acoustic hybrid-order topological
  insulator induced by non-hermiticity and anisotropy},\ }\href
  {https://doi.org/10.1103/PhysRevB.109.134302} {\bibfield  {journal} {\bibinfo
   {journal} {Phys. Rev. B}\ }\textbf {\bibinfo {volume} {109}},\ \bibinfo
  {pages} {134302} (\bibinfo {year} {2024})}\BibitemShut {NoStop}%
\bibitem [{\citenamefont {Zhu}\ \emph {et~al.}(2025)\citenamefont {Zhu},
  \citenamefont {Zheng}, \citenamefont {Palumbo},\ and\ \citenamefont
  {Wang}}]{PhysRevB.111.195107}%
  \BibitemOpen
  \bibfield  {author} {\bibinfo {author} {\bibfnamefont {Y.-Q.}\ \bibnamefont
  {Zhu}}, \bibinfo {author} {\bibfnamefont {Z.}~\bibnamefont {Zheng}}, \bibinfo
  {author} {\bibfnamefont {G.}~\bibnamefont {Palumbo}},\ and\ \bibinfo {author}
  {\bibfnamefont {Z.~D.}\ \bibnamefont {Wang}},\ }\bibfield  {title} {\bibinfo
  {title} {Topological insulators with hybrid-order boundary states},\ }\href
  {https://doi.org/10.1103/PhysRevB.111.195107} {\bibfield  {journal} {\bibinfo
   {journal} {Phys. Rev. B}\ }\textbf {\bibinfo {volume} {111}},\ \bibinfo
  {pages} {195107} (\bibinfo {year} {2025})}\BibitemShut {NoStop}%
\bibitem [{\citenamefont {Li}\ \emph {et~al.}(2018{\natexlab{b}})\citenamefont
  {Li}, \citenamefont {Lee},\ and\ \citenamefont
  {Gong}}]{PhysRevLett.121.036401}%
  \BibitemOpen
  \bibfield  {author} {\bibinfo {author} {\bibfnamefont {L.}~\bibnamefont
  {Li}}, \bibinfo {author} {\bibfnamefont {C.~H.}\ \bibnamefont {Lee}},\ and\
  \bibinfo {author} {\bibfnamefont {J.}~\bibnamefont {Gong}},\ }\bibfield
  {title} {\bibinfo {title} {Realistic {F}loquet {S}emimetal with {E}xotic
  {T}opological {L}inkages between {A}rbitrarily {M}any {N}odal {L}oops},\
  }\href {https://doi.org/10.1103/PhysRevLett.121.036401} {\bibfield  {journal}
  {\bibinfo  {journal} {Phys. Rev. Lett.}\ }\textbf {\bibinfo {volume} {121}},\
  \bibinfo {pages} {036401} (\bibinfo {year} {2018}{\natexlab{b}})}\BibitemShut
  {NoStop}%
\bibitem [{\citenamefont {Hu}\ \emph {et~al.}(2020)\citenamefont {Hu},
  \citenamefont {Huang}, \citenamefont {Zhao},\ and\ \citenamefont
  {Liu}}]{PhysRevLett.124.057001}%
  \BibitemOpen
  \bibfield  {author} {\bibinfo {author} {\bibfnamefont {H.}~\bibnamefont
  {Hu}}, \bibinfo {author} {\bibfnamefont {B.}~\bibnamefont {Huang}}, \bibinfo
  {author} {\bibfnamefont {E.}~\bibnamefont {Zhao}},\ and\ \bibinfo {author}
  {\bibfnamefont {W.~V.}\ \bibnamefont {Liu}},\ }\bibfield  {title} {\bibinfo
  {title} {Dynamical {S}ingularities of {F}loquet {H}igher-{O}rder
  {T}opological {I}nsulators},\ }\href
  {https://doi.org/10.1103/PhysRevLett.124.057001} {\bibfield  {journal}
  {\bibinfo  {journal} {Phys. Rev. Lett.}\ }\textbf {\bibinfo {volume} {124}},\
  \bibinfo {pages} {057001} (\bibinfo {year} {2020})}\BibitemShut {NoStop}%
\bibitem [{\citenamefont {Nag}\ \emph {et~al.}(2021)\citenamefont {Nag},
  \citenamefont {Juri\ifmmode \check{c}\else \v{c}\fi{}i\ifmmode~\acute{c}\else
  \'{c}\fi{}},\ and\ \citenamefont {Roy}}]{PhysRevB.103.115308}%
  \BibitemOpen
  \bibfield  {author} {\bibinfo {author} {\bibfnamefont {T.}~\bibnamefont
  {Nag}}, \bibinfo {author} {\bibfnamefont {V.}~\bibnamefont {Juri\ifmmode
  \check{c}\else \v{c}\fi{}i\ifmmode~\acute{c}\else \'{c}\fi{}}},\ and\
  \bibinfo {author} {\bibfnamefont {B.}~\bibnamefont {Roy}},\ }\bibfield
  {title} {\bibinfo {title} {Hierarchy of higher-order {F}loquet topological
  phases in three dimensions},\ }\href
  {https://doi.org/10.1103/PhysRevB.103.115308} {\bibfield  {journal} {\bibinfo
   {journal} {Phys. Rev. B}\ }\textbf {\bibinfo {volume} {103}},\ \bibinfo
  {pages} {115308} (\bibinfo {year} {2021})}\BibitemShut {NoStop}%
\bibitem [{\citenamefont {Wang}\ \emph {et~al.}(2021)\citenamefont {Wang},
  \citenamefont {Wu},\ and\ \citenamefont {An}}]{PhysRevB.104.205117}%
  \BibitemOpen
  \bibfield  {author} {\bibinfo {author} {\bibfnamefont {B.-Q.}\ \bibnamefont
  {Wang}}, \bibinfo {author} {\bibfnamefont {H.}~\bibnamefont {Wu}},\ and\
  \bibinfo {author} {\bibfnamefont {J.-H.}\ \bibnamefont {An}},\ }\bibfield
  {title} {\bibinfo {title} {Engineering exotic second-order topological
  semimetals by periodic driving},\ }\href
  {https://doi.org/10.1103/PhysRevB.104.205117} {\bibfield  {journal} {\bibinfo
   {journal} {Phys. Rev. B}\ }\textbf {\bibinfo {volume} {104}},\ \bibinfo
  {pages} {205117} (\bibinfo {year} {2021})}\BibitemShut {NoStop}%
\bibitem [{\citenamefont {Ghosh}\ \emph {et~al.}(2022)\citenamefont {Ghosh},
  \citenamefont {Saha},\ and\ \citenamefont {Sengupta}}]{PhysRevB.105.224312}%
  \BibitemOpen
  \bibfield  {author} {\bibinfo {author} {\bibfnamefont {S.}~\bibnamefont
  {Ghosh}}, \bibinfo {author} {\bibfnamefont {K.}~\bibnamefont {Saha}},\ and\
  \bibinfo {author} {\bibfnamefont {K.}~\bibnamefont {Sengupta}},\ }\bibfield
  {title} {\bibinfo {title} {Hinge-mode dynamics of periodically driven
  higher-order weyl semimetals},\ }\href
  {https://doi.org/10.1103/PhysRevB.105.224312} {\bibfield  {journal} {\bibinfo
   {journal} {Phys. Rev. B}\ }\textbf {\bibinfo {volume} {105}},\ \bibinfo
  {pages} {224312} (\bibinfo {year} {2022})}\BibitemShut {NoStop}%
\bibitem [{\citenamefont {Bucciantini}\ \emph {et~al.}(2017)\citenamefont
  {Bucciantini}, \citenamefont {Roy}, \citenamefont {Kitamura},\ and\
  \citenamefont {Oka}}]{PhysRevB.96.041126}%
  \BibitemOpen
  \bibfield  {author} {\bibinfo {author} {\bibfnamefont {L.}~\bibnamefont
  {Bucciantini}}, \bibinfo {author} {\bibfnamefont {S.}~\bibnamefont {Roy}},
  \bibinfo {author} {\bibfnamefont {S.}~\bibnamefont {Kitamura}},\ and\
  \bibinfo {author} {\bibfnamefont {T.}~\bibnamefont {Oka}},\ }\bibfield
  {title} {\bibinfo {title} {Emergent weyl nodes and fermi arcs in a floquet
  weyl semimetal},\ }\href {https://doi.org/10.1103/PhysRevB.96.041126}
  {\bibfield  {journal} {\bibinfo  {journal} {Phys. Rev. B}\ }\textbf {\bibinfo
  {volume} {96}},\ \bibinfo {pages} {041126} (\bibinfo {year}
  {2017})}\BibitemShut {NoStop}%
\bibitem [{\citenamefont {Hübener}\ \emph {et~al.}(2017)\citenamefont
  {Hübener}, \citenamefont {Sentef}, \citenamefont {De~Giovannini},
  \citenamefont {Kemper},\ and\ \citenamefont {Rubio}}]{Hübener.2017}%
  \BibitemOpen
  \bibfield  {author} {\bibinfo {author} {\bibfnamefont {H.}~\bibnamefont
  {Hübener}}, \bibinfo {author} {\bibfnamefont {M.~A.}\ \bibnamefont
  {Sentef}}, \bibinfo {author} {\bibfnamefont {U.}~\bibnamefont
  {De~Giovannini}}, \bibinfo {author} {\bibfnamefont {A.~F.}\ \bibnamefont
  {Kemper}},\ and\ \bibinfo {author} {\bibfnamefont {A.}~\bibnamefont
  {Rubio}},\ }\bibfield  {title} {\bibinfo {title} {Creating stable
  floquet–weyl semimetals by laser-driving of 3d dirac materials},\ }\href
  {https://doi.org/10.1038/ncomms13940} {\bibfield  {journal} {\bibinfo
  {journal} {Nature Communications}\ }\textbf {\bibinfo {volume} {8}},\
  \bibinfo {pages} {13940} (\bibinfo {year} {2017})}\BibitemShut {NoStop}%
\bibitem [{\citenamefont {Nag}\ \emph {et~al.}(2019)\citenamefont {Nag},
  \citenamefont {Slager}, \citenamefont {Higuchi},\ and\ \citenamefont
  {Oka}}]{PhysRevB.100.134301}%
  \BibitemOpen
  \bibfield  {author} {\bibinfo {author} {\bibfnamefont {T.}~\bibnamefont
  {Nag}}, \bibinfo {author} {\bibfnamefont {R.-J.}\ \bibnamefont {Slager}},
  \bibinfo {author} {\bibfnamefont {T.}~\bibnamefont {Higuchi}},\ and\ \bibinfo
  {author} {\bibfnamefont {T.}~\bibnamefont {Oka}},\ }\bibfield  {title}
  {\bibinfo {title} {Dynamical synchronization transition in interacting
  electron systems},\ }\href {https://doi.org/10.1103/PhysRevB.100.134301}
  {\bibfield  {journal} {\bibinfo  {journal} {Phys. Rev. B}\ }\textbf {\bibinfo
  {volume} {100}},\ \bibinfo {pages} {134301} (\bibinfo {year}
  {2019})}\BibitemShut {NoStop}%
\bibitem [{\citenamefont {Wu}\ \emph {et~al.}(2021)\citenamefont {Wu},
  \citenamefont {Wang},\ and\ \citenamefont {An}}]{PhysRevB.103.L041115}%
  \BibitemOpen
  \bibfield  {author} {\bibinfo {author} {\bibfnamefont {H.}~\bibnamefont
  {Wu}}, \bibinfo {author} {\bibfnamefont {B.-Q.}\ \bibnamefont {Wang}},\ and\
  \bibinfo {author} {\bibfnamefont {J.-H.}\ \bibnamefont {An}},\ }\bibfield
  {title} {\bibinfo {title} {Floquet second-order topological insulators in
  non-{H}ermitian systems},\ }\href
  {https://doi.org/10.1103/PhysRevB.103.L041115} {\bibfield  {journal}
  {\bibinfo  {journal} {Phys. Rev. B}\ }\textbf {\bibinfo {volume} {103}},\
  \bibinfo {pages} {L041115} (\bibinfo {year} {2021})}\BibitemShut {NoStop}%
\bibitem [{\citenamefont {Wheeler}\ \emph {et~al.}(2019)\citenamefont
  {Wheeler}, \citenamefont {Wagner},\ and\ \citenamefont
  {Hughes}}]{PhysRevB.100.245135}%
  \BibitemOpen
  \bibfield  {author} {\bibinfo {author} {\bibfnamefont {W.~A.}\ \bibnamefont
  {Wheeler}}, \bibinfo {author} {\bibfnamefont {L.~K.}\ \bibnamefont
  {Wagner}},\ and\ \bibinfo {author} {\bibfnamefont {T.~L.}\ \bibnamefont
  {Hughes}},\ }\bibfield  {title} {\bibinfo {title} {Many-body electric
  multipole operators in extended systems},\ }\href
  {https://doi.org/10.1103/PhysRevB.100.245135} {\bibfield  {journal} {\bibinfo
   {journal} {Phys. Rev. B}\ }\textbf {\bibinfo {volume} {100}},\ \bibinfo
  {pages} {245135} (\bibinfo {year} {2019})}\BibitemShut {NoStop}%
\bibitem [{\citenamefont {Song}\ \emph {et~al.}(2019)\citenamefont {Song},
  \citenamefont {Yao},\ and\ \citenamefont {Wang}}]{PhysRevLett.123.246801}%
  \BibitemOpen
  \bibfield  {author} {\bibinfo {author} {\bibfnamefont {F.}~\bibnamefont
  {Song}}, \bibinfo {author} {\bibfnamefont {S.}~\bibnamefont {Yao}},\ and\
  \bibinfo {author} {\bibfnamefont {Z.}~\bibnamefont {Wang}},\ }\bibfield
  {title} {\bibinfo {title} {Non-{H}ermitian {T}opological {I}nvariants in
  {R}eal {S}pace},\ }\href {https://doi.org/10.1103/PhysRevLett.123.246801}
  {\bibfield  {journal} {\bibinfo  {journal} {Phys. Rev. Lett.}\ }\textbf
  {\bibinfo {volume} {123}},\ \bibinfo {pages} {246801} (\bibinfo {year}
  {2019})}\BibitemShut {NoStop}%
\bibitem [{\citenamefont {Tong}\ \emph {et~al.}(2013)\citenamefont {Tong},
  \citenamefont {An}, \citenamefont {Gong}, \citenamefont {Luo},\ and\
  \citenamefont {Oh}}]{PhysRevB.87.201109}%
  \BibitemOpen
  \bibfield  {author} {\bibinfo {author} {\bibfnamefont {Q.-J.}\ \bibnamefont
  {Tong}}, \bibinfo {author} {\bibfnamefont {J.-H.}\ \bibnamefont {An}},
  \bibinfo {author} {\bibfnamefont {J.}~\bibnamefont {Gong}}, \bibinfo {author}
  {\bibfnamefont {H.-G.}\ \bibnamefont {Luo}},\ and\ \bibinfo {author}
  {\bibfnamefont {C.~H.}\ \bibnamefont {Oh}},\ }\bibfield  {title} {\bibinfo
  {title} {Generating many {M}ajorana modes via periodic driving: A
  superconductor model},\ }\href {https://doi.org/10.1103/PhysRevB.87.201109}
  {\bibfield  {journal} {\bibinfo  {journal} {Phys. Rev. B}\ }\textbf {\bibinfo
  {volume} {87}},\ \bibinfo {pages} {201109} (\bibinfo {year}
  {2013})}\BibitemShut {NoStop}%
\bibitem [{\citenamefont {Else}\ \emph {et~al.}(2016)\citenamefont {Else},
  \citenamefont {Bauer},\ and\ \citenamefont {Nayak}}]{PhysRevLett.117.090402}%
  \BibitemOpen
  \bibfield  {author} {\bibinfo {author} {\bibfnamefont {D.~V.}\ \bibnamefont
  {Else}}, \bibinfo {author} {\bibfnamefont {B.}~\bibnamefont {Bauer}},\ and\
  \bibinfo {author} {\bibfnamefont {C.}~\bibnamefont {Nayak}},\ }\bibfield
  {title} {\bibinfo {title} {Floquet {T}ime {C}rystals},\ }\href
  {https://doi.org/10.1103/PhysRevLett.117.090402} {\bibfield  {journal}
  {\bibinfo  {journal} {Phys. Rev. Lett.}\ }\textbf {\bibinfo {volume} {117}},\
  \bibinfo {pages} {090402} (\bibinfo {year} {2016})}\BibitemShut {NoStop}%
\bibitem [{\citenamefont {Yao}\ \emph {et~al.}(2017)\citenamefont {Yao},
  \citenamefont {Potter}, \citenamefont {Potirniche},\ and\ \citenamefont
  {Vishwanath}}]{PhysRevLett.118.030401}%
  \BibitemOpen
  \bibfield  {author} {\bibinfo {author} {\bibfnamefont {N.~Y.}\ \bibnamefont
  {Yao}}, \bibinfo {author} {\bibfnamefont {A.~C.}\ \bibnamefont {Potter}},
  \bibinfo {author} {\bibfnamefont {I.-D.}\ \bibnamefont {Potirniche}},\ and\
  \bibinfo {author} {\bibfnamefont {A.}~\bibnamefont {Vishwanath}},\ }\bibfield
   {title} {\bibinfo {title} {Discrete {T}ime {C}rystals: {R}igidity,
  {C}riticality, and {R}ealizations},\ }\href
  {https://doi.org/10.1103/PhysRevLett.118.030401} {\bibfield  {journal}
  {\bibinfo  {journal} {Phys. Rev. Lett.}\ }\textbf {\bibinfo {volume} {118}},\
  \bibinfo {pages} {030401} (\bibinfo {year} {2017})}\BibitemShut {NoStop}%
\bibitem [{\citenamefont {Sambe}(1973)}]{PhysRevA.7.2203}%
  \BibitemOpen
  \bibfield  {author} {\bibinfo {author} {\bibfnamefont {H.}~\bibnamefont
  {Sambe}},\ }\bibfield  {title} {\bibinfo {title} {Steady {S}tates and
  {Q}uasienergies of a {Q}uantum-{M}echanical {S}ystem in an {O}scillating
  {F}ield},\ }\href {https://doi.org/10.1103/PhysRevA.7.2203} {\bibfield
  {journal} {\bibinfo  {journal} {Phys. Rev. A}\ }\textbf {\bibinfo {volume}
  {7}},\ \bibinfo {pages} {2203} (\bibinfo {year} {1973})}\BibitemShut
  {NoStop}%
\bibitem [{\citenamefont {Chen}\ \emph {et~al.}(2015)\citenamefont {Chen},
  \citenamefont {An}, \citenamefont {Luo}, \citenamefont {Sun},\ and\
  \citenamefont {Oh}}]{PhysRevA.91.052122}%
  \BibitemOpen
  \bibfield  {author} {\bibinfo {author} {\bibfnamefont {C.}~\bibnamefont
  {Chen}}, \bibinfo {author} {\bibfnamefont {J.-H.}\ \bibnamefont {An}},
  \bibinfo {author} {\bibfnamefont {H.-G.}\ \bibnamefont {Luo}}, \bibinfo
  {author} {\bibfnamefont {C.~P.}\ \bibnamefont {Sun}},\ and\ \bibinfo {author}
  {\bibfnamefont {C.~H.}\ \bibnamefont {Oh}},\ }\bibfield  {title} {\bibinfo
  {title} {Floquet control of quantum dissipation in spin chains},\ }\href
  {https://doi.org/10.1103/PhysRevA.91.052122} {\bibfield  {journal} {\bibinfo
  {journal} {Phys. Rev. A}\ }\textbf {\bibinfo {volume} {91}},\ \bibinfo
  {pages} {052122} (\bibinfo {year} {2015})}\BibitemShut {NoStop}%
\bibitem [{\citenamefont {Asb\'oth}\ \emph {et~al.}(2014)\citenamefont
  {Asb\'oth}, \citenamefont {Tarasinski},\ and\ \citenamefont
  {Delplace}}]{PhysRevB.90.125143}%
  \BibitemOpen
  \bibfield  {author} {\bibinfo {author} {\bibfnamefont {J.~K.}\ \bibnamefont
  {Asb\'oth}}, \bibinfo {author} {\bibfnamefont {B.}~\bibnamefont
  {Tarasinski}},\ and\ \bibinfo {author} {\bibfnamefont {P.}~\bibnamefont
  {Delplace}},\ }\bibfield  {title} {\bibinfo {title} {Chiral symmetry and
  bulk-boundary correspondence in periodically driven one-dimensional
  systems},\ }\href {https://doi.org/10.1103/PhysRevB.90.125143} {\bibfield
  {journal} {\bibinfo  {journal} {Phys. Rev. B}\ }\textbf {\bibinfo {volume}
  {90}},\ \bibinfo {pages} {125143} (\bibinfo {year} {2014})}\BibitemShut
  {NoStop}%
\bibitem [{\citenamefont {Kobayashi}\ \emph {et~al.}(2017)\citenamefont
  {Kobayashi}, \citenamefont {Yamakawa}, \citenamefont {Yamakage},
  \citenamefont {Inohara}, \citenamefont {Okamoto},\ and\ \citenamefont
  {Tanaka}}]{PhysRevB.95.245208}%
  \BibitemOpen
  \bibfield  {author} {\bibinfo {author} {\bibfnamefont {S.}~\bibnamefont
  {Kobayashi}}, \bibinfo {author} {\bibfnamefont {Y.}~\bibnamefont {Yamakawa}},
  \bibinfo {author} {\bibfnamefont {A.}~\bibnamefont {Yamakage}}, \bibinfo
  {author} {\bibfnamefont {T.}~\bibnamefont {Inohara}}, \bibinfo {author}
  {\bibfnamefont {Y.}~\bibnamefont {Okamoto}},\ and\ \bibinfo {author}
  {\bibfnamefont {Y.}~\bibnamefont {Tanaka}},\ }\bibfield  {title} {\bibinfo
  {title} {Crossing-line-node semimetals: {G}eneral theory and application to
  rare-earth trihydrides},\ }\href {https://doi.org/10.1103/PhysRevB.95.245208}
  {\bibfield  {journal} {\bibinfo  {journal} {Phys. Rev. B}\ }\textbf {\bibinfo
  {volume} {95}},\ \bibinfo {pages} {245208} (\bibinfo {year}
  {2017})}\BibitemShut {NoStop}%
\bibitem [{\citenamefont {Wei}\ \emph {et~al.}(2024)\citenamefont {Wei},
  \citenamefont {He}, \citenamefont {Yan}, \citenamefont {He},\ and\
  \citenamefont {Jia}}]{Wei_2024}%
  \BibitemOpen
  \bibfield  {author} {\bibinfo {author} {\bibfnamefont {Q.}~\bibnamefont
  {Wei}}, \bibinfo {author} {\bibfnamefont {A.-L.}\ \bibnamefont {He}},
  \bibinfo {author} {\bibfnamefont {M.}~\bibnamefont {Yan}}, \bibinfo {author}
  {\bibfnamefont {H.}~\bibnamefont {He}},\ and\ \bibinfo {author}
  {\bibfnamefont {S.}~\bibnamefont {Jia}},\ }\bibfield  {title} {\bibinfo
  {title} {Hybrid-order weyl semimetal and its acoustic realizations},\ }\href
  {https://doi.org/10.1088/1361-6463/ad6b35} {\bibfield  {journal} {\bibinfo
  {journal} {Journal of Physics D: Applied Physics}\ }\textbf {\bibinfo
  {volume} {57}},\ \bibinfo {pages} {465303} (\bibinfo {year}
  {2024})}\BibitemShut {NoStop}%
\bibitem [{\citenamefont {Bzdušek}\ \emph {et~al.}(2016)\citenamefont
  {Bzdušek}, \citenamefont {Wu}, \citenamefont {Rüegg}, \citenamefont
  {Sigrist},\ and\ \citenamefont {Soluyanov}}]{Bzdusek.2016}%
  \BibitemOpen
  \bibfield  {author} {\bibinfo {author} {\bibfnamefont {T.}~\bibnamefont
  {Bzdušek}}, \bibinfo {author} {\bibfnamefont {Q.}~\bibnamefont {Wu}},
  \bibinfo {author} {\bibfnamefont {A.}~\bibnamefont {Rüegg}}, \bibinfo
  {author} {\bibfnamefont {M.}~\bibnamefont {Sigrist}},\ and\ \bibinfo {author}
  {\bibfnamefont {A.~A.}\ \bibnamefont {Soluyanov}},\ }\bibfield  {title}
  {\bibinfo {title} {Nodal-chain metals},\ }\href
  {https://doi.org/10.1038/nature19099} {\bibfield  {journal} {\bibinfo
  {journal} {Nature}\ }\textbf {\bibinfo {volume} {538}},\ \bibinfo {pages}
  {75} (\bibinfo {year} {2016})}\BibitemShut {NoStop}%
\bibitem [{\citenamefont {Wang}\ \emph {et~al.}(2018)\citenamefont {Wang},
  \citenamefont {Nie}, \citenamefont {Weng}, \citenamefont {Kawazoe},\ and\
  \citenamefont {Chen}}]{PhysRevLett.120.026402}%
  \BibitemOpen
  \bibfield  {author} {\bibinfo {author} {\bibfnamefont {J.-T.}\ \bibnamefont
  {Wang}}, \bibinfo {author} {\bibfnamefont {S.}~\bibnamefont {Nie}}, \bibinfo
  {author} {\bibfnamefont {H.}~\bibnamefont {Weng}}, \bibinfo {author}
  {\bibfnamefont {Y.}~\bibnamefont {Kawazoe}},\ and\ \bibinfo {author}
  {\bibfnamefont {C.}~\bibnamefont {Chen}},\ }\bibfield  {title} {\bibinfo
  {title} {Topological {N}odal-{N}et {S}emimetal in a {G}raphene {N}etwork
  {S}tructure},\ }\href {https://doi.org/10.1103/PhysRevLett.120.026402}
  {\bibfield  {journal} {\bibinfo  {journal} {Phys. Rev. Lett.}\ }\textbf
  {\bibinfo {volume} {120}},\ \bibinfo {pages} {026402} (\bibinfo {year}
  {2018})}\BibitemShut {NoStop}%
\bibitem [{\citenamefont {Li}\ \emph {et~al.}(2017)\citenamefont {Li},
  \citenamefont {Yin}, \citenamefont {Chen},\ and\ \citenamefont
  {Ara\'ujo}}]{PhysRevB.95.121107}%
  \BibitemOpen
  \bibfield  {author} {\bibinfo {author} {\bibfnamefont {L.}~\bibnamefont
  {Li}}, \bibinfo {author} {\bibfnamefont {C.}~\bibnamefont {Yin}}, \bibinfo
  {author} {\bibfnamefont {S.}~\bibnamefont {Chen}},\ and\ \bibinfo {author}
  {\bibfnamefont {M.~A.~N.}\ \bibnamefont {Ara\'ujo}},\ }\bibfield  {title}
  {\bibinfo {title} {Chiral topological insulating phases from
  three-dimensional nodal loop semimetals},\ }\href
  {https://doi.org/10.1103/PhysRevB.95.121107} {\bibfield  {journal} {\bibinfo
  {journal} {Phys. Rev. B}\ }\textbf {\bibinfo {volume} {95}},\ \bibinfo
  {pages} {121107} (\bibinfo {year} {2017})}\BibitemShut {NoStop}%
\bibitem [{\citenamefont {Wei}\ \emph {et~al.}(2021)\citenamefont {Wei},
  \citenamefont {Zhang}, \citenamefont {Deng}, \citenamefont {Lu},
  \citenamefont {Huang}, \citenamefont {Yan}, \citenamefont {Chen},
  \citenamefont {Liu},\ and\ \citenamefont {Jia}}]{WeiQiang2021}%
  \BibitemOpen
  \bibfield  {author} {\bibinfo {author} {\bibfnamefont {Q.}~\bibnamefont
  {Wei}}, \bibinfo {author} {\bibfnamefont {X.}~\bibnamefont {Zhang}}, \bibinfo
  {author} {\bibfnamefont {W.}~\bibnamefont {Deng}}, \bibinfo {author}
  {\bibfnamefont {J.}~\bibnamefont {Lu}}, \bibinfo {author} {\bibfnamefont
  {X.}~\bibnamefont {Huang}}, \bibinfo {author} {\bibfnamefont
  {M.}~\bibnamefont {Yan}}, \bibinfo {author} {\bibfnamefont {G.}~\bibnamefont
  {Chen}}, \bibinfo {author} {\bibfnamefont {Z.}~\bibnamefont {Liu}},\ and\
  \bibinfo {author} {\bibfnamefont {S.}~\bibnamefont {Jia}},\ }\bibfield
  {title} {\bibinfo {title} {Higher-order topological semimetal in acoustic
  crystals},\ }\href {https://doi.org/10.1038/s41563-021-00933-4} {\bibfield
  {journal} {\bibinfo  {journal} {Nature Materials}\ }\textbf {\bibinfo
  {volume} {20}},\ \bibinfo {pages} {812} (\bibinfo {year} {2021})}\BibitemShut
  {NoStop}%
\bibitem [{\citenamefont {Qiu}\ \emph {et~al.}(2021)\citenamefont {Qiu},
  \citenamefont {Xiao}, \citenamefont {Zhang},\ and\ \citenamefont
  {Qiu}}]{PhysRevLett.127.146601}%
  \BibitemOpen
  \bibfield  {author} {\bibinfo {author} {\bibfnamefont {H.}~\bibnamefont
  {Qiu}}, \bibinfo {author} {\bibfnamefont {M.}~\bibnamefont {Xiao}}, \bibinfo
  {author} {\bibfnamefont {F.}~\bibnamefont {Zhang}},\ and\ \bibinfo {author}
  {\bibfnamefont {C.}~\bibnamefont {Qiu}},\ }\bibfield  {title} {\bibinfo
  {title} {Higher-{O}rder {D}irac {S}onic {C}rystals},\ }\href
  {https://doi.org/10.1103/PhysRevLett.127.146601} {\bibfield  {journal}
  {\bibinfo  {journal} {Phys. Rev. Lett.}\ }\textbf {\bibinfo {volume} {127}},\
  \bibinfo {pages} {146601} (\bibinfo {year} {2021})}\BibitemShut {NoStop}%
\bibitem [{\citenamefont {Xia}\ \emph {et~al.}(2022)\citenamefont {Xia},
  \citenamefont {Lai}, \citenamefont {Sun}, \citenamefont {He},\ and\
  \citenamefont {Chen}}]{PhysRevLett.128.115701}%
  \BibitemOpen
  \bibfield  {author} {\bibinfo {author} {\bibfnamefont {C.-H.}\ \bibnamefont
  {Xia}}, \bibinfo {author} {\bibfnamefont {H.-S.}\ \bibnamefont {Lai}},
  \bibinfo {author} {\bibfnamefont {X.-C.}\ \bibnamefont {Sun}}, \bibinfo
  {author} {\bibfnamefont {C.}~\bibnamefont {He}},\ and\ \bibinfo {author}
  {\bibfnamefont {Y.-F.}\ \bibnamefont {Chen}},\ }\bibfield  {title} {\bibinfo
  {title} {Experimental {D}emonstration of {B}ulk-{H}inge {C}orrespondence in a
  {T}hree-{D}imensional {T}opological {D}irac {A}coustic {C}rystal},\ }\href
  {https://doi.org/10.1103/PhysRevLett.128.115701} {\bibfield  {journal}
  {\bibinfo  {journal} {Phys. Rev. Lett.}\ }\textbf {\bibinfo {volume} {128}},\
  \bibinfo {pages} {115701} (\bibinfo {year} {2022})}\BibitemShut {NoStop}%
\bibitem [{\citenamefont {Pu}\ \emph {et~al.}(2023)\citenamefont {Pu},
  \citenamefont {He}, \citenamefont {Luo}, \citenamefont {Ma}, \citenamefont
  {Ye}, \citenamefont {Ke},\ and\ \citenamefont
  {Liu}}]{PhysRevLett.130.116103}%
  \BibitemOpen
  \bibfield  {author} {\bibinfo {author} {\bibfnamefont {Z.}~\bibnamefont
  {Pu}}, \bibinfo {author} {\bibfnamefont {H.}~\bibnamefont {He}}, \bibinfo
  {author} {\bibfnamefont {L.}~\bibnamefont {Luo}}, \bibinfo {author}
  {\bibfnamefont {Q.}~\bibnamefont {Ma}}, \bibinfo {author} {\bibfnamefont
  {L.}~\bibnamefont {Ye}}, \bibinfo {author} {\bibfnamefont {M.}~\bibnamefont
  {Ke}},\ and\ \bibinfo {author} {\bibfnamefont {Z.}~\bibnamefont {Liu}},\
  }\bibfield  {title} {\bibinfo {title} {Acoustic {H}igher-{O}rder {W}eyl
  {S}emimetal with {B}ound {H}inge {S}tates in the {C}ontinuum},\ }\href
  {https://doi.org/10.1103/PhysRevLett.130.116103} {\bibfield  {journal}
  {\bibinfo  {journal} {Phys. Rev. Lett.}\ }\textbf {\bibinfo {volume} {130}},\
  \bibinfo {pages} {116103} (\bibinfo {year} {2023})}\BibitemShut {NoStop}%
\bibitem [{\citenamefont {Wang}\ \emph {et~al.}(2022)\citenamefont {Wang},
  \citenamefont {Liu}, \citenamefont {Teo}, \citenamefont {Wang}, \citenamefont
  {Xue},\ and\ \citenamefont {Zhang}}]{PhysRevB.105.L060101}%
  \BibitemOpen
  \bibfield  {author} {\bibinfo {author} {\bibfnamefont {Z.}~\bibnamefont
  {Wang}}, \bibinfo {author} {\bibfnamefont {D.}~\bibnamefont {Liu}}, \bibinfo
  {author} {\bibfnamefont {H.~T.}\ \bibnamefont {Teo}}, \bibinfo {author}
  {\bibfnamefont {Q.}~\bibnamefont {Wang}}, \bibinfo {author} {\bibfnamefont
  {H.}~\bibnamefont {Xue}},\ and\ \bibinfo {author} {\bibfnamefont
  {B.}~\bibnamefont {Zhang}},\ }\bibfield  {title} {\bibinfo {title}
  {Higher-order {D}irac semimetal in a photonic crystal},\ }\href
  {https://doi.org/10.1103/PhysRevB.105.L060101} {\bibfield  {journal}
  {\bibinfo  {journal} {Phys. Rev. B}\ }\textbf {\bibinfo {volume} {105}},\
  \bibinfo {pages} {L060101} (\bibinfo {year} {2022})}\BibitemShut {NoStop}%
\bibitem [{\citenamefont {Pan}\ \emph {et~al.}(2023)\citenamefont {Pan},
  \citenamefont {Cui}, \citenamefont {Chen}, \citenamefont {Chen},
  \citenamefont {Zhang}, \citenamefont {Ren}, \citenamefont {Han},
  \citenamefont {Li}, \citenamefont {Li}, \citenamefont {Yu}, \citenamefont
  {Chen},\ and\ \citenamefont {Yang}}]{PanYuang2023}%
  \BibitemOpen
  \bibfield  {author} {\bibinfo {author} {\bibfnamefont {Y.}~\bibnamefont
  {Pan}}, \bibinfo {author} {\bibfnamefont {C.}~\bibnamefont {Cui}}, \bibinfo
  {author} {\bibfnamefont {Q.}~\bibnamefont {Chen}}, \bibinfo {author}
  {\bibfnamefont {F.}~\bibnamefont {Chen}}, \bibinfo {author} {\bibfnamefont
  {L.}~\bibnamefont {Zhang}}, \bibinfo {author} {\bibfnamefont
  {Y.}~\bibnamefont {Ren}}, \bibinfo {author} {\bibfnamefont {N.}~\bibnamefont
  {Han}}, \bibinfo {author} {\bibfnamefont {W.}~\bibnamefont {Li}}, \bibinfo
  {author} {\bibfnamefont {X.}~\bibnamefont {Li}}, \bibinfo {author}
  {\bibfnamefont {Z.-M.}\ \bibnamefont {Yu}}, \bibinfo {author} {\bibfnamefont
  {H.}~\bibnamefont {Chen}},\ and\ \bibinfo {author} {\bibfnamefont
  {Y.}~\bibnamefont {Yang}},\ }\bibfield  {title} {\bibinfo {title} {Real
  higher-order {W}eyl photonic crystal},\ }\href
  {https://doi.org/10.1038/s41467-023-42457-2} {\bibfield  {journal} {\bibinfo
  {journal} {Nature Communications}\ }\textbf {\bibinfo {volume} {14}},\
  \bibinfo {pages} {6636} (\bibinfo {year} {2023})}\BibitemShut {NoStop}%
\bibitem [{\citenamefont {Song}\ \emph {et~al.}(2022)\citenamefont {Song},
  \citenamefont {Yang}, \citenamefont {Cao},\ and\ \citenamefont
  {Yan}}]{SongLingling2022}%
  \BibitemOpen
  \bibfield  {author} {\bibinfo {author} {\bibfnamefont {L.}~\bibnamefont
  {Song}}, \bibinfo {author} {\bibfnamefont {H.}~\bibnamefont {Yang}}, \bibinfo
  {author} {\bibfnamefont {Y.}~\bibnamefont {Cao}},\ and\ \bibinfo {author}
  {\bibfnamefont {P.}~\bibnamefont {Yan}},\ }\bibfield  {title} {\bibinfo
  {title} {Square-root higher-order {W}eyl semimetals},\ }\href
  {https://doi.org/10.1038/s41467-022-33306-9} {\bibfield  {journal} {\bibinfo
  {journal} {Nature Communications}\ }\textbf {\bibinfo {volume} {13}},\
  \bibinfo {pages} {5601} (\bibinfo {year} {2022})}\BibitemShut {NoStop}%
\bibitem [{\citenamefont {Xiang}\ \emph {et~al.}(2024)\citenamefont {Xiang},
  \citenamefont {Peng}, \citenamefont {Gao}, \citenamefont {Wu}, \citenamefont
  {Wu}, \citenamefont {Chen}, \citenamefont {Ni},\ and\ \citenamefont
  {Zhu}}]{PhysRevLett.132.197202}%
  \BibitemOpen
  \bibfield  {author} {\bibinfo {author} {\bibfnamefont {X.}~\bibnamefont
  {Xiang}}, \bibinfo {author} {\bibfnamefont {Y.-G.}\ \bibnamefont {Peng}},
  \bibinfo {author} {\bibfnamefont {F.}~\bibnamefont {Gao}}, \bibinfo {author}
  {\bibfnamefont {X.}~\bibnamefont {Wu}}, \bibinfo {author} {\bibfnamefont
  {P.}~\bibnamefont {Wu}}, \bibinfo {author} {\bibfnamefont {Z.}~\bibnamefont
  {Chen}}, \bibinfo {author} {\bibfnamefont {X.}~\bibnamefont {Ni}},\ and\
  \bibinfo {author} {\bibfnamefont {X.-F.}\ \bibnamefont {Zhu}},\ }\bibfield
  {title} {\bibinfo {title} {Demonstration of {A}coustic {H}igher-{O}rder
  {T}opological {S}tiefel-{W}hitney {S}emimetal},\ }\href
  {https://doi.org/10.1103/PhysRevLett.132.197202} {\bibfield  {journal}
  {\bibinfo  {journal} {Phys. Rev. Lett.}\ }\textbf {\bibinfo {volume} {132}},\
  \bibinfo {pages} {197202} (\bibinfo {year} {2024})}\BibitemShut {NoStop}%
\bibitem [{\citenamefont {Xue}\ \emph {et~al.}(2023)\citenamefont {Xue},
  \citenamefont {Chen}, \citenamefont {Cheng}, \citenamefont {Dai},
  \citenamefont {Long}, \citenamefont {Zhao},\ and\ \citenamefont
  {Zhang}}]{XueHaoran2023}%
  \BibitemOpen
  \bibfield  {author} {\bibinfo {author} {\bibfnamefont {H.}~\bibnamefont
  {Xue}}, \bibinfo {author} {\bibfnamefont {Z.~Y.}\ \bibnamefont {Chen}},
  \bibinfo {author} {\bibfnamefont {Z.}~\bibnamefont {Cheng}}, \bibinfo
  {author} {\bibfnamefont {J.~X.}\ \bibnamefont {Dai}}, \bibinfo {author}
  {\bibfnamefont {Y.}~\bibnamefont {Long}}, \bibinfo {author} {\bibfnamefont
  {Y.~X.}\ \bibnamefont {Zhao}},\ and\ \bibinfo {author} {\bibfnamefont
  {B.}~\bibnamefont {Zhang}},\ }\bibfield  {title} {\bibinfo {title}
  {Stiefel-whitney topological charges in a three-dimensional acoustic
  nodal-line crystal},\ }\href {https://doi.org/10.1038/s41467-023-40252-7}
  {\bibfield  {journal} {\bibinfo  {journal} {Nature Communications}\ }\textbf
  {\bibinfo {volume} {14}},\ \bibinfo {pages} {4563} (\bibinfo {year}
  {2023})}\BibitemShut {NoStop}%
\bibitem [{\citenamefont {Ma}\ \emph {et~al.}(2024)\citenamefont {Ma},
  \citenamefont {Pu}, \citenamefont {Ye}, \citenamefont {Lu}, \citenamefont
  {Huang}, \citenamefont {Ke}, \citenamefont {He}, \citenamefont {Deng},\ and\
  \citenamefont {Liu}}]{PhysRevLett.132.066601}%
  \BibitemOpen
  \bibfield  {author} {\bibinfo {author} {\bibfnamefont {Q.}~\bibnamefont
  {Ma}}, \bibinfo {author} {\bibfnamefont {Z.}~\bibnamefont {Pu}}, \bibinfo
  {author} {\bibfnamefont {L.}~\bibnamefont {Ye}}, \bibinfo {author}
  {\bibfnamefont {J.}~\bibnamefont {Lu}}, \bibinfo {author} {\bibfnamefont
  {X.}~\bibnamefont {Huang}}, \bibinfo {author} {\bibfnamefont
  {M.}~\bibnamefont {Ke}}, \bibinfo {author} {\bibfnamefont {H.}~\bibnamefont
  {He}}, \bibinfo {author} {\bibfnamefont {W.}~\bibnamefont {Deng}},\ and\
  \bibinfo {author} {\bibfnamefont {Z.}~\bibnamefont {Liu}},\ }\bibfield
  {title} {\bibinfo {title} {Observation of higher-order nodal-line semimetal
  in phononic crystals},\ }\href
  {https://doi.org/10.1103/PhysRevLett.132.066601} {\bibfield  {journal}
  {\bibinfo  {journal} {Phys. Rev. Lett.}\ }\textbf {\bibinfo {volume} {132}},\
  \bibinfo {pages} {066601} (\bibinfo {year} {2024})}\BibitemShut {NoStop}%
\bibitem [{\citenamefont {Neupane}\ \emph {et~al.}(2014)\citenamefont
  {Neupane}, \citenamefont {Xu}, \citenamefont {Sankar}, \citenamefont
  {Alidoust}, \citenamefont {Bian}, \citenamefont {Liu}, \citenamefont
  {Belopolski}, \citenamefont {Chang}, \citenamefont {Jeng}, \citenamefont
  {Lin}, \citenamefont {Bansil}, \citenamefont {Chou},\ and\ \citenamefont
  {Hasan}}]{Neupane2014}%
  \BibitemOpen
  \bibfield  {author} {\bibinfo {author} {\bibfnamefont {M.}~\bibnamefont
  {Neupane}}, \bibinfo {author} {\bibfnamefont {S.-Y.}\ \bibnamefont {Xu}},
  \bibinfo {author} {\bibfnamefont {R.}~\bibnamefont {Sankar}}, \bibinfo
  {author} {\bibfnamefont {N.}~\bibnamefont {Alidoust}}, \bibinfo {author}
  {\bibfnamefont {G.}~\bibnamefont {Bian}}, \bibinfo {author} {\bibfnamefont
  {C.}~\bibnamefont {Liu}}, \bibinfo {author} {\bibfnamefont {I.}~\bibnamefont
  {Belopolski}}, \bibinfo {author} {\bibfnamefont {T.-R.}\ \bibnamefont
  {Chang}}, \bibinfo {author} {\bibfnamefont {H.-T.}\ \bibnamefont {Jeng}},
  \bibinfo {author} {\bibfnamefont {H.}~\bibnamefont {Lin}}, \bibinfo {author}
  {\bibfnamefont {A.}~\bibnamefont {Bansil}}, \bibinfo {author} {\bibfnamefont
  {F.}~\bibnamefont {Chou}},\ and\ \bibinfo {author} {\bibfnamefont {M.~Z.}\
  \bibnamefont {Hasan}},\ }\bibfield  {title} {\bibinfo {title} {Observation of
  a three-dimensional topological dirac semimetal phase in high-mobility
  ${C}d_{3}{A}s_{2}$},\ }\href {https://doi.org/10.1038/ncomms4786} {\bibfield
  {journal} {\bibinfo  {journal} {Nature Communications}\ }\textbf {\bibinfo
  {volume} {5}},\ \bibinfo {pages} {3786} (\bibinfo {year} {2014})}\BibitemShut
  {NoStop}%
\bibitem [{\citenamefont {Schoop}\ \emph {et~al.}(2016)\citenamefont {Schoop},
  \citenamefont {Ali}, \citenamefont {Straßer}, \citenamefont {Topp},
  \citenamefont {Varykhalov}, \citenamefont {Marchenko}, \citenamefont
  {Duppel}, \citenamefont {Parkin}, \citenamefont {Lotsch},\ and\ \citenamefont
  {Ast}}]{Schoop2016}%
  \BibitemOpen
  \bibfield  {author} {\bibinfo {author} {\bibfnamefont {L.~M.}\ \bibnamefont
  {Schoop}}, \bibinfo {author} {\bibfnamefont {M.~N.}\ \bibnamefont {Ali}},
  \bibinfo {author} {\bibfnamefont {C.}~\bibnamefont {Straßer}}, \bibinfo
  {author} {\bibfnamefont {A.}~\bibnamefont {Topp}}, \bibinfo {author}
  {\bibfnamefont {A.}~\bibnamefont {Varykhalov}}, \bibinfo {author}
  {\bibfnamefont {D.}~\bibnamefont {Marchenko}}, \bibinfo {author}
  {\bibfnamefont {V.}~\bibnamefont {Duppel}}, \bibinfo {author} {\bibfnamefont
  {S.~S.~P.}\ \bibnamefont {Parkin}}, \bibinfo {author} {\bibfnamefont {B.~V.}\
  \bibnamefont {Lotsch}},\ and\ \bibinfo {author} {\bibfnamefont {C.~R.}\
  \bibnamefont {Ast}},\ }\bibfield  {title} {\bibinfo {title} {Dirac cone
  protected by non-symmorphic symmetry and three-dimensional dirac line node in
  ${Z}r{S}i{S}$},\ }\href {https://doi.org/10.1038/ncomms11696} {\bibfield
  {journal} {\bibinfo  {journal} {Nature Communications}\ }\textbf {\bibinfo
  {volume} {7}},\ \bibinfo {pages} {11696} (\bibinfo {year}
  {2016})}\BibitemShut {NoStop}%
\bibitem [{\citenamefont {Cheng}\ \emph {et~al.}(2019)\citenamefont {Cheng},
  \citenamefont {Pan}, \citenamefont {Wang}, \citenamefont {Zhang},
  \citenamefont {Yu}, \citenamefont {Gover}, \citenamefont {Zhang},
  \citenamefont {Li}, \citenamefont {Zhou},\ and\ \citenamefont
  {Zhu}}]{PhysRevLett.122.173901}%
  \BibitemOpen
  \bibfield  {author} {\bibinfo {author} {\bibfnamefont {Q.}~\bibnamefont
  {Cheng}}, \bibinfo {author} {\bibfnamefont {Y.}~\bibnamefont {Pan}}, \bibinfo
  {author} {\bibfnamefont {H.}~\bibnamefont {Wang}}, \bibinfo {author}
  {\bibfnamefont {C.}~\bibnamefont {Zhang}}, \bibinfo {author} {\bibfnamefont
  {D.}~\bibnamefont {Yu}}, \bibinfo {author} {\bibfnamefont {A.}~\bibnamefont
  {Gover}}, \bibinfo {author} {\bibfnamefont {H.}~\bibnamefont {Zhang}},
  \bibinfo {author} {\bibfnamefont {T.}~\bibnamefont {Li}}, \bibinfo {author}
  {\bibfnamefont {L.}~\bibnamefont {Zhou}},\ and\ \bibinfo {author}
  {\bibfnamefont {S.}~\bibnamefont {Zhu}},\ }\bibfield  {title} {\bibinfo
  {title} {Observation of {A}nomalous $\ensuremath{\pi}$ {M}odes in {P}hotonic
  {F}loquet {E}ngineering},\ }\href
  {https://doi.org/10.1103/PhysRevLett.122.173901} {\bibfield  {journal}
  {\bibinfo  {journal} {Phys. Rev. Lett.}\ }\textbf {\bibinfo {volume} {122}},\
  \bibinfo {pages} {173901} (\bibinfo {year} {2019})}\BibitemShut {NoStop}%
\bibitem [{\citenamefont {Maczewsky}\ \emph {et~al.}(2017)\citenamefont
  {Maczewsky}, \citenamefont {Zeuner}, \citenamefont {Nolte},\ and\
  \citenamefont {Szameit}}]{Maczewsky.2017}%
  \BibitemOpen
  \bibfield  {author} {\bibinfo {author} {\bibfnamefont {L.~J.}\ \bibnamefont
  {Maczewsky}}, \bibinfo {author} {\bibfnamefont {J.~M.}\ \bibnamefont
  {Zeuner}}, \bibinfo {author} {\bibfnamefont {S.}~\bibnamefont {Nolte}},\ and\
  \bibinfo {author} {\bibfnamefont {A.}~\bibnamefont {Szameit}},\ }\bibfield
  {title} {\bibinfo {title} {Observation of photonic anomalous {F}loquet
  topological insulators},\ }\href {https://doi.org/10.1038/ncomms13756}
  {\bibfield  {journal} {\bibinfo  {journal} {Nature Communications}\ }\textbf
  {\bibinfo {volume} {8}},\ \bibinfo {pages} {13756} (\bibinfo {year}
  {2017})}\BibitemShut {NoStop}%
\bibitem [{\citenamefont {Mukherjee}\ \emph {et~al.}(2017)\citenamefont
  {Mukherjee}, \citenamefont {Spracklen}, \citenamefont {Valiente},
  \citenamefont {Andersson}, \citenamefont {Öhberg}, \citenamefont {Goldman},\
  and\ \citenamefont {Thomson}}]{Mukherjee.2017}%
  \BibitemOpen
  \bibfield  {author} {\bibinfo {author} {\bibfnamefont {S.}~\bibnamefont
  {Mukherjee}}, \bibinfo {author} {\bibfnamefont {A.}~\bibnamefont
  {Spracklen}}, \bibinfo {author} {\bibfnamefont {M.}~\bibnamefont {Valiente}},
  \bibinfo {author} {\bibfnamefont {E.}~\bibnamefont {Andersson}}, \bibinfo
  {author} {\bibfnamefont {P.}~\bibnamefont {Öhberg}}, \bibinfo {author}
  {\bibfnamefont {N.}~\bibnamefont {Goldman}},\ and\ \bibinfo {author}
  {\bibfnamefont {R.~R.}\ \bibnamefont {Thomson}},\ }\bibfield  {title}
  {\bibinfo {title} {Experimental observation of anomalous topological edge
  modes in a slowly driven photonic lattice},\ }\href
  {https://doi.org/10.1038/ncomms13918} {\bibfield  {journal} {\bibinfo
  {journal} {Nature Communications}\ }\textbf {\bibinfo {volume} {8}},\
  \bibinfo {pages} {13918} (\bibinfo {year} {2017})}\BibitemShut {NoStop}%
\bibitem [{\citenamefont {Wintersperger}\ \emph {et~al.}(2020)\citenamefont
  {Wintersperger}, \citenamefont {Braun}, \citenamefont {Ünal}, \citenamefont
  {Eckardt}, \citenamefont {Liberto}, \citenamefont {Goldman}, \citenamefont
  {Bloch},\ and\ \citenamefont {Aidelsburger}}]{Wintersperger.2010}%
  \BibitemOpen
  \bibfield  {author} {\bibinfo {author} {\bibfnamefont {K.}~\bibnamefont
  {Wintersperger}}, \bibinfo {author} {\bibfnamefont {C.}~\bibnamefont
  {Braun}}, \bibinfo {author} {\bibfnamefont {F.~N.}\ \bibnamefont {Ünal}},
  \bibinfo {author} {\bibfnamefont {A.}~\bibnamefont {Eckardt}}, \bibinfo
  {author} {\bibfnamefont {M.~D.}\ \bibnamefont {Liberto}}, \bibinfo {author}
  {\bibfnamefont {N.}~\bibnamefont {Goldman}}, \bibinfo {author} {\bibfnamefont
  {I.}~\bibnamefont {Bloch}},\ and\ \bibinfo {author} {\bibfnamefont
  {M.}~\bibnamefont {Aidelsburger}},\ }\bibfield  {title} {\bibinfo {title}
  {Realization of an anomalous {F}loquet topological system with ultracold
  atoms},\ }\href {https://doi.org/10.1038/s41567-020-0949-y} {\bibfield
  {journal} {\bibinfo  {journal} {Nature Physics}\ }\textbf {\bibinfo {volume}
  {16}},\ \bibinfo {pages} {1058} (\bibinfo {year} {2020})}\BibitemShut
  {NoStop}%
\bibitem [{\citenamefont {Roushan}\ \emph {et~al.}(2017)\citenamefont
  {Roushan}, \citenamefont {Neill}, \citenamefont {Megrant}, \citenamefont
  {Chen}, \citenamefont {Babbush}, \citenamefont {Barends}, \citenamefont
  {Campbell}, \citenamefont {Chen}, \citenamefont {Chiaro}, \citenamefont
  {Dunsworth}, \citenamefont {Fowler}, \citenamefont {Jeffrey}, \citenamefont
  {Kelly}, \citenamefont {Lucero}, \citenamefont {Mutus}, \citenamefont
  {O’Malley}, \citenamefont {Neeley}, \citenamefont {Quintana}, \citenamefont
  {Sank}, \citenamefont {Vainsencher}, \citenamefont {Wenner}, \citenamefont
  {White}, \citenamefont {Kapit}, \citenamefont {Neven},\ and\ \citenamefont
  {Martinis}}]{Roushan.P.2017}%
  \BibitemOpen
  \bibfield  {author} {\bibinfo {author} {\bibfnamefont {P.}~\bibnamefont
  {Roushan}}, \bibinfo {author} {\bibfnamefont {C.}~\bibnamefont {Neill}},
  \bibinfo {author} {\bibfnamefont {A.}~\bibnamefont {Megrant}}, \bibinfo
  {author} {\bibfnamefont {Y.}~\bibnamefont {Chen}}, \bibinfo {author}
  {\bibfnamefont {R.}~\bibnamefont {Babbush}}, \bibinfo {author} {\bibfnamefont
  {R.}~\bibnamefont {Barends}}, \bibinfo {author} {\bibfnamefont
  {B.}~\bibnamefont {Campbell}}, \bibinfo {author} {\bibfnamefont
  {Z.}~\bibnamefont {Chen}}, \bibinfo {author} {\bibfnamefont {B.}~\bibnamefont
  {Chiaro}}, \bibinfo {author} {\bibfnamefont {A.}~\bibnamefont {Dunsworth}},
  \bibinfo {author} {\bibfnamefont {A.}~\bibnamefont {Fowler}}, \bibinfo
  {author} {\bibfnamefont {E.}~\bibnamefont {Jeffrey}}, \bibinfo {author}
  {\bibfnamefont {J.}~\bibnamefont {Kelly}}, \bibinfo {author} {\bibfnamefont
  {E.}~\bibnamefont {Lucero}}, \bibinfo {author} {\bibfnamefont
  {J.}~\bibnamefont {Mutus}}, \bibinfo {author} {\bibfnamefont {P.~J.~J.}\
  \bibnamefont {O’Malley}}, \bibinfo {author} {\bibfnamefont
  {M.}~\bibnamefont {Neeley}}, \bibinfo {author} {\bibfnamefont
  {C.}~\bibnamefont {Quintana}}, \bibinfo {author} {\bibfnamefont
  {D.}~\bibnamefont {Sank}}, \bibinfo {author} {\bibfnamefont {A.}~\bibnamefont
  {Vainsencher}}, \bibinfo {author} {\bibfnamefont {J.}~\bibnamefont {Wenner}},
  \bibinfo {author} {\bibfnamefont {T.}~\bibnamefont {White}}, \bibinfo
  {author} {\bibfnamefont {E.}~\bibnamefont {Kapit}}, \bibinfo {author}
  {\bibfnamefont {H.}~\bibnamefont {Neven}},\ and\ \bibinfo {author}
  {\bibfnamefont {J.}~\bibnamefont {Martinis}},\ }\bibfield  {title} {\bibinfo
  {title} {Chiral ground-state currents of interacting photons in a synthetic
  magnetic field},\ }\href {https://doi.org/10.1038/nphys3930} {\bibfield
  {journal} {\bibinfo  {journal} {Nature Physics}\ }\textbf {\bibinfo {volume}
  {13}},\ \bibinfo {pages} {146} (\bibinfo {year} {2017})}\BibitemShut
  {NoStop}%
\bibitem [{\citenamefont {Mahmood}\ \emph {et~al.}(2016)\citenamefont
  {Mahmood}, \citenamefont {Chan}, \citenamefont {Alpichshev}, \citenamefont
  {Gardner}, \citenamefont {Lee}, \citenamefont {Lee},\ and\ \citenamefont
  {Gedik}}]{Fahad.2016}%
  \BibitemOpen
  \bibfield  {author} {\bibinfo {author} {\bibfnamefont {F.}~\bibnamefont
  {Mahmood}}, \bibinfo {author} {\bibfnamefont {C.-K.}\ \bibnamefont {Chan}},
  \bibinfo {author} {\bibfnamefont {Z.}~\bibnamefont {Alpichshev}}, \bibinfo
  {author} {\bibfnamefont {D.}~\bibnamefont {Gardner}}, \bibinfo {author}
  {\bibfnamefont {Y.}~\bibnamefont {Lee}}, \bibinfo {author} {\bibfnamefont
  {P.~A.}\ \bibnamefont {Lee}},\ and\ \bibinfo {author} {\bibfnamefont
  {N.}~\bibnamefont {Gedik}},\ }\bibfield  {title} {\bibinfo {title} {Selective
  scattering between {F}loquet–{B}loch and {V}olkov states in a topological
  insulator},\ }\href {https://doi.org/10.1038/nphys3609} {\bibfield  {journal}
  {\bibinfo  {journal} {Nature Physics}\ }\textbf {\bibinfo {volume} {12}},\
  \bibinfo {pages} {306} (\bibinfo {year} {2016})}\BibitemShut {NoStop}%
\bibitem [{\citenamefont {McIver}\ \emph {et~al.}(2020)\citenamefont {McIver},
  \citenamefont {Schulte}, \citenamefont {Stein}, \citenamefont {Matsuyama},
  \citenamefont {Jotzu}, \citenamefont {Meier},\ and\ \citenamefont
  {Cavalleri}}]{McIver.2020}%
  \BibitemOpen
  \bibfield  {author} {\bibinfo {author} {\bibfnamefont {J.~W.}\ \bibnamefont
  {McIver}}, \bibinfo {author} {\bibfnamefont {B.}~\bibnamefont {Schulte}},
  \bibinfo {author} {\bibfnamefont {F.~U.}\ \bibnamefont {Stein}}, \bibinfo
  {author} {\bibfnamefont {T.}~\bibnamefont {Matsuyama}}, \bibinfo {author}
  {\bibfnamefont {G.}~\bibnamefont {Jotzu}}, \bibinfo {author} {\bibfnamefont
  {G.}~\bibnamefont {Meier}},\ and\ \bibinfo {author} {\bibfnamefont
  {A.}~\bibnamefont {Cavalleri}},\ }\bibfield  {title} {\bibinfo {title}
  {Light-induced anomalous {H}all effect in graphene},\ }\href
  {https://doi.org/10.1038/s41567-019-0698-y} {\bibfield  {journal} {\bibinfo
  {journal} {Nature Physics}\ }\textbf {\bibinfo {volume} {16}},\ \bibinfo
  {pages} {38} (\bibinfo {year} {2020})}\BibitemShut {NoStop}%
\bibitem [{\citenamefont {Dabiri}\ and\ \citenamefont
  {Cheraghchi}(2023)}]{10.1063/5.0150118}%
  \BibitemOpen
  \bibfield  {author} {\bibinfo {author} {\bibfnamefont {S.~S.}\ \bibnamefont
  {Dabiri}}\ and\ \bibinfo {author} {\bibfnamefont {H.}~\bibnamefont
  {Cheraghchi}},\ }\bibfield  {title} {\bibinfo {title} {Electric circuit
  simulation of floquet topological insulators in fourier space},\ }\href
  {https://doi.org/10.1063/5.0150118} {\bibfield  {journal} {\bibinfo
  {journal} {Journal of Applied Physics}\ }\textbf {\bibinfo {volume} {134}},\
  \bibinfo {pages} {084303} (\bibinfo {year} {2023})}\BibitemShut {NoStop}%
\bibitem [{\citenamefont {Chitsazi}\ \emph {et~al.}(2017)\citenamefont
  {Chitsazi}, \citenamefont {Li}, \citenamefont {Ellis},\ and\ \citenamefont
  {Kottos}}]{PhysRevLett.119.093901}%
  \BibitemOpen
  \bibfield  {author} {\bibinfo {author} {\bibfnamefont {M.}~\bibnamefont
  {Chitsazi}}, \bibinfo {author} {\bibfnamefont {H.}~\bibnamefont {Li}},
  \bibinfo {author} {\bibfnamefont {F.~M.}\ \bibnamefont {Ellis}},\ and\
  \bibinfo {author} {\bibfnamefont {T.}~\bibnamefont {Kottos}},\ }\bibfield
  {title} {\bibinfo {title} {Experimental realization of floquet
  $\mathcal{P}\mathcal{T}$-symmetric systems},\ }\href
  {https://doi.org/10.1103/PhysRevLett.119.093901} {\bibfield  {journal}
  {\bibinfo  {journal} {Phys. Rev. Lett.}\ }\textbf {\bibinfo {volume} {119}},\
  \bibinfo {pages} {093901} (\bibinfo {year} {2017})}\BibitemShut {NoStop}%
\end{thebibliography}%

\end{document}